\documentclass[aps,twocolumn,showkeys,superscriptaddress]{revtex4-1}

\usepackage{graphicx}
\usepackage{ifpdf}
\usepackage{bm}	
\usepackage{latexsym}
\usepackage{color,soul}
\usepackage{pstricks}
\usepackage{figsize}
\usepackage{float}
\usepackage{color}
\usepackage{amssymb}
\usepackage{hyperref}
\usepackage{amsmath}
\usepackage{epstopdf}
\usepackage{verbatim}
\usepackage{bbold}

\newcommand{\p}{\partial}
\newcommand{\vphi}{\varphi}

\begin{document}

%\preprint{}

\title{Fermion parity measurement and control in Majorana circuit quantum electrodynamics}


\author{Konstantin Yavilberg}
\affiliation{Department of Physics, Ben-Gurion University of the
Negev, Be’er-Sheva 84105, Israel}

\author{Eran Ginossar}
\affiliation{Advanced Technology Institute and Department of Physics, University of Surrey, Guildford GU2 7XH, United Kingdom}

\author{Eytan Grosfeld}
\affiliation{Department of Physics, Ben-Gurion University of the
Negev, Be’er-Sheva 84105, Israel}

\begin{abstract} 
We investigate the quantum electrodynamics of a device based on a topological superconducting circuit embedded in a microwave resonator. The device stores its quantum information in coherent superpositions of fermion parity states originating from Majorana fermion hybridization. This generates a highly isolated qubit whose coherence time could be greatly enhanced. We extend the conventional semiclassical method and obtain analytical derivations for strong transmon-photon coupling. Using this formalism, we develop protocols to initialize, control, and measure the parity states. We show that, remarkably, the parity eigenvalue can be detected via dispersive shifts of the optical cavity in the strong-coupling regime and its state can be coherently manipulated via a second-order sideband transition.
\end{abstract}
%\keywords{}

\maketitle
\section{Introduction}
Advances occurring over the past decade have given rise to a new generation of single-qubit solid-state architectures which hold the promise of compatibility with existing electronics and fabrication techniques. Among these devices, superconducting circuit processors based on the transmon qubit \cite{koch_charge-insensitive_2007,schreier2008suppressing,houck2008controlling} have shown great potential in terms of coherent control, measurement, and scalability. These provide a unique opportunity to study fundamental quantum phenomena in engineered macroscopic two-level systems which are controlled by coherent microwave photons. Of particular interest is the study of hybrid devices where a microscopic or a mesoscopic system is embedded within the superconducting circuit. The properties of the constituent devices can contribute to the optimization of the qubit, including the processes related to its preparation, manipulation, and readout, to its coherence properties, and to its prospects for scaling up.

The integration of Majorana fermions \cite{read2000paired,kitaev2007unpaired} into the superconducting circuit architecture \cite{hassler2011top,bonderson2011topological,van2011coulomb,
schmidt2013majorana,pekker2013proposal,blais_majorana_2013,cottet_majorana_squeezing_2013,
GinossarGrosfeld2014,PhysRevB.91.085406} can potentially lead to improved qubits, with the ultimate goal being the realization of a topologically protected information storage and high-coherence processing device \cite{hassler2011top,bonderson2011topological}. The recently introduced Majorana transmon (MT) \citep{GinossarGrosfeld2014} sacrifices full topological protection by directly exploiting a weak interaction between two neighboring Majorana fermions, but gains a highly anharmonic spectrum, composed of well-separated nearly degenerate doublets, originating from parity states hybridization. In order to facilitate both detection and control, the circuit should be embedded within a microwave resonator \cite{PhysRevA.69.062320,PhysRevA.74.042318,PhysRevB.86.100506,	Majer2007,PhysRevLett.75.3788,:/content/aip/journal/jap/104/11/10.1063/1.3010859} where the  strong interaction with the cavity field will provide the means for coherent qubit control and readout. In the proposed device, the lowest doublet of states is analogous to the familiar ion hyperfine qubit \cite{BenhelmIonsQIP2008} and does not couple directly to the cavity and the radiative environment owing to both a small matrix element and its small frequency. The consequence is an  increased qubit lifetime and therefore potentially very high readout and control fidelities.

\begin{figure}[h]
	\centering
		\includegraphics[scale=0.21]{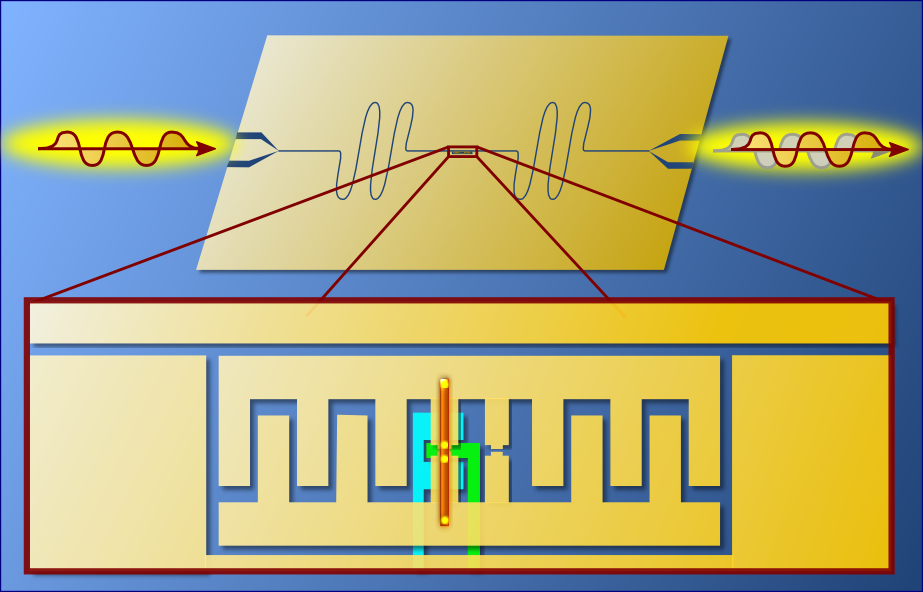}
	\caption{(Color online) Majorana-transmon circuit in a cavity. A topological superconductor (orange) bridges a Josephson junction nucleating Majorana fermions (yellow). Control gates (light blue and green) are used to drive the topological state and control the Majorana coupling. Together with its superconducting islands, this device is embedded inside a coplanar transmission-line resonator. Microwave pulses on the input port allow logical gates realizations, while the output port can be used for homodyne or heterodyne detection of the fermion parity state.}
	\label{fig:mcqed-setup}
\end{figure}

In this paper we develop the hybrid circuit quantum electrodynamics of the MT system strongly coupled to a single-mode electromagnetic field (see Fig.~\ref{fig:mcqed-setup}). First, we use a semiclassical approximation to obtain the eigenstates and spectrum of the device. We then use these to find expressions for the dipole matrix elements, demonstrating that the doublet forming the logical qubit is coupled to a higher doublet which can serve as a control. When the microwave transitions of the MT device are detuned from the cavity resonance, we show that a dispersive interaction arises between the fermionic parity, the transmon oscillator, and the cavity photon degrees of freedom. We use this regime to propose a scheme for measuring the state of the qubit, which is revealed via a fine structure in the cavity dispersive frequency shifts that are, remarkably, sensitive to the different fermionic parities associated with the two levels. In addition, we discuss protocols for (i) qubit cooling and (ii) implementing a single qubit rotation, demonstrating qubit control. Together, these provide the minimal ingredients required for establishing the relevance of the device for quantum information processing \cite{PhysRevA.52.3457}.

The measurement and control scheme that we develop here is based on an indirect coupling of photons to the parity doublet mediated through transitions to higher transmon levels. Another approach relies on the presence of a dipole coupling between the doublet states \cite{PhysRevB.91.085406}, which is, however, extremely small in our model when taken to the charge-noise resilient transmon regime. Other proposed qubit measurement and control schemes require electronic shuttling of Majorana fermions between superconducting islands using a series of depletion gates, relying on the adiabaticity of this process \cite{hassler2011top,blais_majorana_2013}.

\section{Results}
\subsection{Eigenstates of the Majorana-transmon device}
The system we study consists of a Josephson junction capacitively coupled to a gate creating an offset charge $n_g$ between the superconducting islands. A nano wire which can support Majorana fermions \cite{lutchyn2010majorana,oreg2010helical} is placed along the junction. Another realization could be based on the recent discovery of Majorana fermions in a chain of magnetic impurities \cite{Nadj-Perge02102014}, where here the chain should cross the Josephson junction. The zero-energy Dirac fermions composed of these Majorana fermions allow the relative number of Cooper pairs, $\hat{n}=-i\p_\vphi$, to admit both integer and half-integer values \cite{fu2010electron}. Such a set-up can be described by the model Hamiltonian $H=H_T \mathbb{1}+H_M\tau_x$, where $\tau_i$ ($i=x,y,z$) are Pauli matrices operating in parity space.
The first term in $H$, containing the transmon Hamiltonian $H_T=-4E_C\p_\vphi^2-E_J\cos(\vphi)$, has the eigenstates $|k\rangle_e = \left[f_k(\vphi),0 \right]^T$, $|k\rangle_o =\left[0,g_k(\vphi)\right]^T$, which correspond to an even and odd fermionic parity. These wave functions obey the boundary conditions $f_k(\vphi+2\pi)=e^{- 2\pi i n_g}f_k(\vphi)$, $g_k(\vphi+2\pi)=e^{-2\pi i (n_g+1/2)}g_k(\vphi)$, which were chosen to ensure the correct quantization of the charge. Since the transmon operates in the regime $E_J/E_C\gg 1$, there is a close resemblance to an anharmonic oscillator which suggests using  a semiclassical derivation of its eigenvalues and eigenstates if the parity-related boundary conditions could be accounted for (see the Appendix).
The second term $H_M= E_M \cos(\vphi/2)$ describes the interaction between the adjacent Majorana fermions \cite{van2011coulomb,GinossarGrosfeld2014,Eur.Phys.37.3} and creates a condensate ``parity flip'' accompanying a single electron tunneling process across the junction. In the setup, we consider $E_M,E_C$, and $E_J$ are all independent energy scales. In the following, we assume $E_M\ll E_C$.

We proceed with the diagonalization of the Hamiltonian which consists of approximately independent transmon bands, with two interlaced parity subbands split due to the Majorana interaction. We define the overlap between the odd and even states due to $H_M$,
${m_{kk'} = E_M \int_{-\infty}^\infty \Psi_k(\vphi) \Psi_{k'}(\vphi)\cos(\vphi/2)d\vphi}$,
where we used the harmonic-oscillator states $\Psi_k(\vphi)$ as an approximation for $f_k(\vphi)$ and $g_k(\vphi)$. The intraband coupling is independent of the specific band and it is dominated by the interaction energy, $m_{kk} \simeq E_M$ to a leading order. The coupling between the bands decreases as $m_{kk'}\sim\left(E_C/E_J\right)^{|k-k'|/2}$, for even $|k-k'|$, and vanishes for odd $|k-k'|$. By neglecting terms of order $\sim\sqrt{E_C/E_J}$ and higher, the Hamiltonian matrix takes a block diagonal form $H = \bigoplus_{k=0}^\infty H^{(k)}$. In the $|k\rangle_e$, $|k\rangle_o$ basis, these blocks are given by
\begin{equation}
\label{eq:mt-hamiltonian block}
	 H^{(k)} = 
	 \begin{pmatrix}
	 	\epsilon_k +t_k \cos(2\pi n_g) & E_M \\
	 	E_M & \epsilon_k -t_k \cos(2\pi n_g)
	 \end{pmatrix},
\end{equation}
where $\epsilon_k$ are the energies of the harmonic oscillator with a first-order anharmonic correction and $ t_k=(-1)^{k+1}\frac{2^{4(k+1)}E_C}{k!}\sqrt{\frac{2}{\pi}}\left(\frac{E_J}{2E_C}\right)^{\frac{k}{2}+\frac{3}{4}}\exp\left(-\sqrt{\frac{8E_J}{E_C}}\right)$ is the transmon dispersion (see the Appendix). The matrix can be diagonalized by a rotation around the $y$ axis, $U=e^{i\eta_k\tau_y}$, $H^{(k)}\to U H^{(k)} U^\dag$, where $\eta_k=\frac{(-1)^{k+1}}{2}\text{atan2}\left[E_M,(-1)^{k+1}t_k \cos(2\pi n_g)\right]$ and $\text{atan2}(y,x)\equiv 2 \tan^{-1}\left[y/(\sqrt{x^2+y^2}+x)\right]$ is the quadrant-dependent arctangent. The eigenvectors of Eq.~(\ref{eq:mt-hamiltonian block}) are
\begin{equation}
\label{eq:mt-mt eigenstates}
\begin{aligned}
	|k,-\rangle & = \cos(\eta_k)|k\rangle_e + \sin(\eta_k)|k\rangle_o, \\
	|k,+\rangle & = -\sin(\eta_k)|k\rangle_e + \cos(\eta_k)|k\rangle_o, \\
\end{aligned}
\end{equation}
and have corresponding eigenvalues
\begin{equation}
\label{eq:mt-mt eigenvalues}
	E_{k,s} = \epsilon_k +s (-1)^k\sqrt{E_M^2 + t_k^2 \cos^2(2\pi n_g)}.
\end{equation}
where the quantum number $s=\pm$ corresponds to the rotated parity according to Eq.~(\ref{eq:mt-mt eigenstates}). It can be seen that the MT eigenvalues are further flattened by the presence of $E_M$, improving on the charge-noise resilience of the transmon \cite{koch_charge-insensitive_2007}. The eigenstates are superpositions of transmon wavefunctions with only a parametric dependence on $n_g$. As we now discuss, this dependence leads to interference effects in the dipole transitions, and thus can be exploited to control the qubit using a cavity.

\subsection{Generalized Jaynes-Cummings model}
We now couple the MT to a single mode of a quantized electromagnetic field in the microwave range, confined within a cavity consisting of a transmission-line resonator \cite{Majer2007,Wallraff2004}. The full quantum description of this system is given by the Rabi Hamiltonian \cite{koch_charge-insensitive_2007,PhysRevA.69.062320} ($\hbar=1$) ${H_{R} = \left(H_T+\omega_c a^\dag a\right)\mathbb{1} + H_M \tau_x + \hat{\mathcal{G}}\left(a^\dag+a\right),}$
where $\omega_c$ is the frequency of the photons created (annihilated) by the operator $a^\dag$ ($a$). The interaction between the MT and the cavity is achieved via the dipole coupling $\hat{\mathcal{G}}=g\left(i\partial_\vphi-n_g\right)\mathbb{1}$, where $g=2e\beta\mathcal{E}_\text{rms} d$ is the dipole coupling strength, $d$ is the distance between the superconducting islands, $\mathcal{E}_\text{rms}$ is the rms field at the ground state of the resonator, and $\beta$ is the ratio between the gate capacitance and the capacitance of the junction. In the harmonic-oscillator approximation, the parity of the wave functions ensures that the dipole transitions are non zero only between neighboring bands. Projecting $H_R$ on the eigenstates given in Eq.~(\ref{eq:mt-mt eigenstates}) and using the rotating-wave approximation (RWA), the Hamiltonian takes the generalized Jaynes-Cummings (JC) form
\begin{equation}
\label{eq:jc-hamiltonian mt basis}
\begin{aligned}
	H_{JC} &= \sum_{ks}E_{k,s}|k,s\rangle \langle k,s|+ \omega_c a^\dag a\\
	  +&\left(\sum_{kss'}\mathcal{G}_{k,s;k+1,s'}|k,s\rangle \langle k+1,s'|a^\dag + \text{h.c.}\right),
\end{aligned}
\end{equation}
where $\mathcal{G}_{k,s;k+1,s'}=\langle k,s |\hat{\mathcal{G}}|k+1,s'\rangle$.
We proceed to analyze the Hamiltonian, given by Eq.~(\ref{eq:jc-hamiltonian mt basis}), in the combined qubit-cavity basis $|k,s;N\rangle$, with associated energies $E_{k,s;N} = E_{k,s}+N\omega_c$, where $N$ is the photon number in the cavity.
The doublet states $|0,s;0\rangle$, which will serve as the qubit, decouple from the cavity interaction and, within the semiclassical approximation, intradoublet transitions vanish. For a nonzero photon number in the cavity, the qubit states are dressed to first order in the dipole coupling, only with the first excited transmon level $|1,s;N\rangle$, according to the $\ell$th excitation sector of the Hamiltonian (here, $\ell=k+N\geq 1$):
\begin{equation}
\label{eq:jc-N photons block}
	H_\ell=
	\begin{pmatrix}
		E_{1,+;\ell-1} & 0 & -\sqrt{\ell}\mathcal{G}_x^* & \sqrt{\ell}\mathcal{G}_o^* \\
		0 & E_{1,-;\ell-1} & \sqrt{\ell}\mathcal{G}_o^* & \sqrt{\ell}\mathcal{G}_x^* \\
		-\sqrt{\ell}\mathcal{G}_x & \sqrt{\ell}\mathcal{G}_o & E_{0,-;\ell} & 0 \\
		\sqrt{\ell}\mathcal{G}_o & \sqrt{\ell}\mathcal{G}_x & 0 & E_{0,+;\ell} 
	\end{pmatrix}.
\end{equation}
The relevant dipole matrix elements are
\begin{figure}[t]
	\centering
	\includegraphics[width=0.8\linewidth]{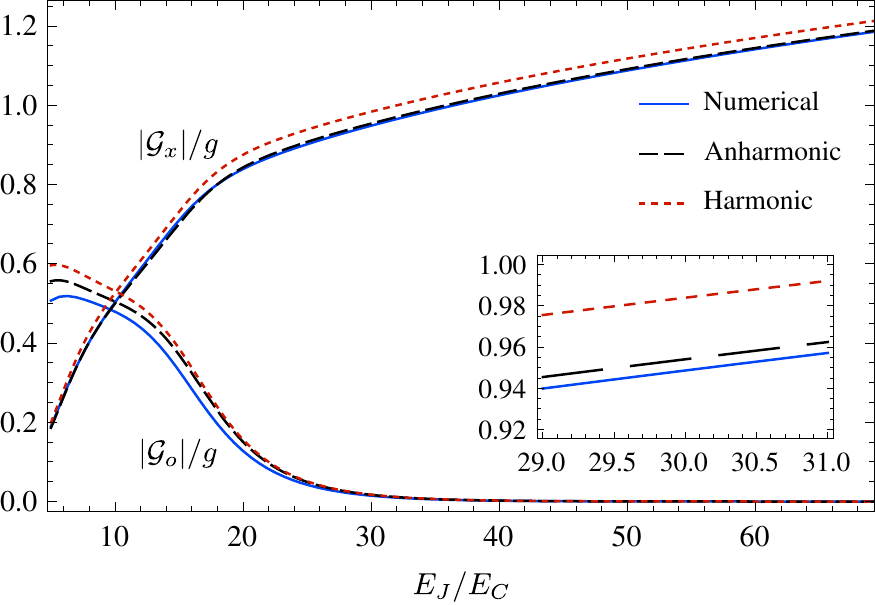}
	\caption{(Color online) Comparison of the WKB result for the dipole matrix elements with numerics. Dipole transition strength is plotted vs $E_J/E_C$ for $n_g=0.2$ and $E_M=0.01 E_C$. The $\mathcal{G}_o$ coupling corresponds to transitions which conserve $s=\pm$ of Eq.~(\ref{eq:mt-mt eigenvalues}), while the $\mathcal{G}_x$ coupling describes transitions which flip $s$ [see also Fig.~\ref{fig:cooling-control}(c)]. Within the transmon range, the anharmonic approximation is more accurate but should eventually coincide with the harmonic approximation for $E_J\to\infty$.}
	\label{fig:transmon-setup}
\end{figure}
\begin{equation}
\begin{aligned}
	\label{eq:dipole}
	\mathcal{G}_o=\mathcal{G}_T \cos(\eta_1-\eta_0),\quad \mathcal{G}_x= \mathcal{G}_T \sin(\eta_1-\eta_0),
\end{aligned}
\end{equation}
where $\mathcal{G}_T$ is the dipole transition associated with the transmon,
\begin{equation}
\label{eq:dipole-transmon value}
	\mathcal{G}_T = ig\int_{-\infty}^\infty \Psi_0(\vphi) \Psi_1'(\vphi)d\vphi = ig\left(\frac{E_J}{32E_C}\right)^{1/4}.
\end{equation}
These results are compared against numerics in Fig.~\ref{fig:transmon-setup}. As demonstrated in the figure, the addition of anharmonic corrections to $\mathcal{G}_T$ further improves the agreement with numerics, so one can replace $\mathcal{G}_T\to \mathcal{G}^{\text{ah}}_T=\mathcal{G}_T\left(1-\sqrt{\frac{E_C}{32 E_J}}+\frac{15E_C}{256 E_J}\right)$ in Eq.~(\ref{eq:dipole}).

\subsection{Spectroscopy and parity state detection}
By coupling the MT to a high-{\it{Q}} superconducting resonator and tuning it to the strong dispersive regime 
\cite{1439774,:/content/aip/journal/apl/100/19/10.1063/1.4710520,Wallraff2004,Niemczyk2010,PhysRevA.80.033846,PhysRevA.81.042311}
, it would be possible to probe the telling features of the spectrum \cite{GinossarGrosfeld2014} and to verify its dependence on the parameters $n_g$ and $E_J$ that were discussed in the previous sections. Experimentally these would involve a homodyne measurement setup for detecting changes in the complex amplitude of a coherent microwave tone which is transmitted at the resonance of the cavity, while spectroscopic pulses are driving the four MT microwave resonances. When a spectroscopic pulse excites the qubit, the transmission through the cavity resonance is diminished, indicating the transition frequency. The pulse can be emitted from a dedicated transmission line terminating in the vicinity of the qubit \cite{PhysRevLett.104.100504}. This measurement only requires the usual dispersive interaction between the cavity occupation and the transmon state \cite{PhysRevLett.94.123602}. Extending the applicability of the dispersive measurement to the parity state requires an additional dependence of the resonator frequency on the fermionic parity.

\begin{figure}[h]
        \centering
			\includegraphics[width=0.8\linewidth]{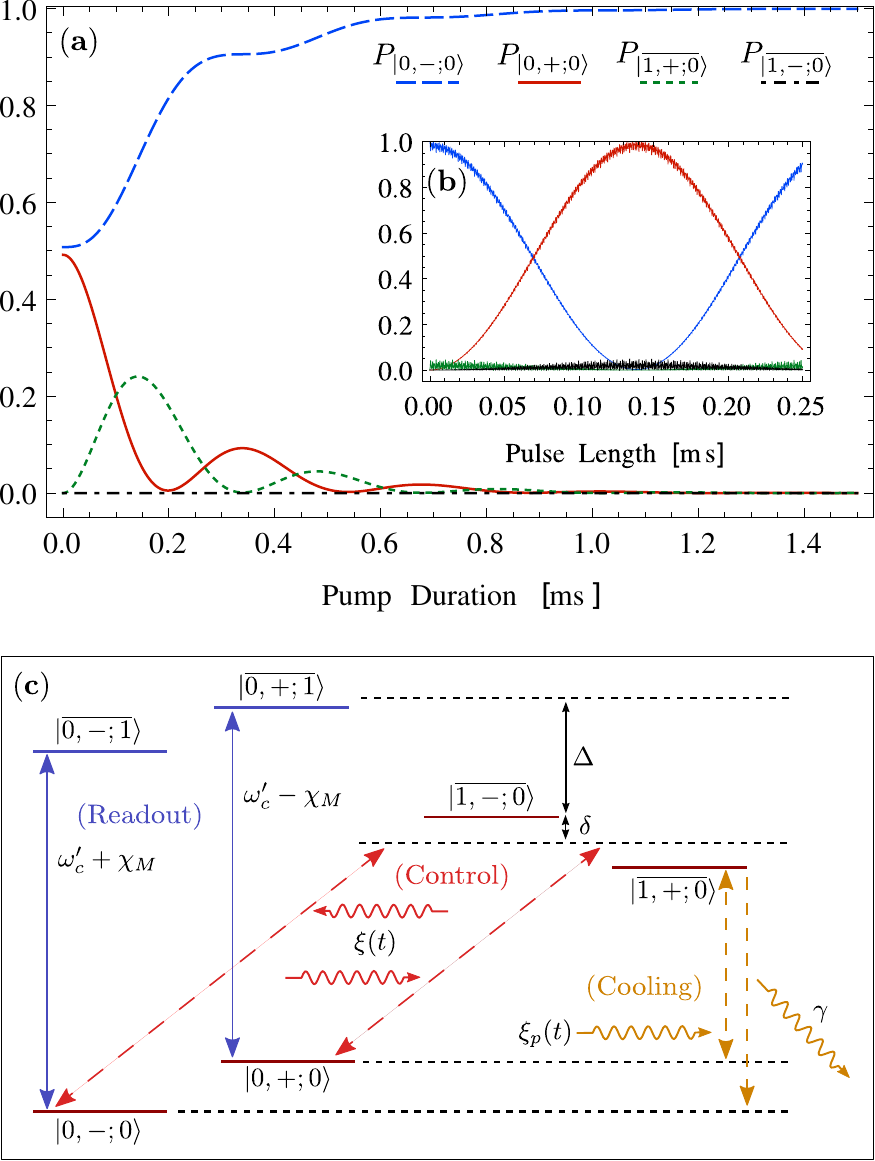}
        \caption{(Color online) Qubit initialization and coherent control in the strong-dispersive regime. (a) Qubit initialization: Numerical calculation of the four levels population during the cooling process. A microwave pump with amplitude $\xi_p/h=10^{-3}$ GHz and frequency  $\omega/2\pi = 5.34\,$GHz is admitted to the system with a preliminary temperature of $0.03\;$K. We take the decay rate as $\gamma/2\pi= 10^{-5}\,$GHz. $E_M=0.025 E_C$, $E_J=25E_C$, $E_C/h=0.4\,$GHz, $n_g=0$, $g/\Delta=0.3$. (b) Coherent control: The system prepared in its ground state and driven via two coherent microwave tones using a rectangular pulse with amplitude $\xi/h= 8 \times 10^{-4}$ GHz and detuning $\delta=E_M$, resulting in transfer between the lower levels with a minimal higher-level population. (c) Level diagram: 
The qubit-cavity interaction (left) produces a shifted resonance frequency $\omega_c' = \omega_c + \chi_T$ accompanied by a parity-based shift $\pm\chi_M$. A process of coherent control (middle) in the double $\Lambda$ system is preceded by the cooling process (right).}
        \label{fig:cooling-control}
\end{figure}

To see how these interactions arise here we return to the system Hamiltonian, given by Eq.~(\ref{eq:jc-N photons block}), and assume that $E_M$ dominates over the transmon dispersion. 
%The closed part of the hamiltonian (\emph{i.e.}, without drive) in the $\ell$'th excitation sector 
We diagonalize to get $H_{\ell} = \omega_c \ell-\sigma_z\sqrt{(\Delta/2)^2+|\mathcal{G}_T|^2\ell+h_M}$ with $h_M=E_M^2+E_M \tau_z\sqrt{\Delta^2+4|\mathcal{G}_x|^2 \ell}$, where $\tau_z$ and $\sigma_z$ operate in the rotated parity ($s=\pm$) and the dressed transmon ($k=0,1$) degrees of freedom, respectively; and $\Delta=\omega_c-(E_{1,+}-E_{0,-})$ is the cavity-transmon detuning.
We next express $H_{\ell}$ with the quantum numbers of the eigenstates ($N,\sigma_z,\tau_z$) and in the dispersive regime, and for small $N$ we expand it in the small parameters $|\mathcal{G}_{o,x}|/\Delta$, $E_M/\Delta$. The resulting diagonal matrix in the basis $|\sigma_z,\tau_z,N\rangle$ represents the effective Hamiltonian
\begin{equation}
\label{eq:jc-effective hamiltonian}
\begin{aligned}
	H_\text{eff} &= \left(\frac{\omega_c-\Delta}{2}-E_M\tau_z\right)\sigma_z+\frac{\mathbb{1}+\sigma_z}{2}\left(-\chi_T+\chi_M\tau_z\right) \\
	& +a^\dag a\left(\omega_c - \chi_T\sigma_z+\chi_M\sigma_z\tau_z\right).
\end{aligned}
\end{equation}
The dispersive shift $\chi_T=|\mathcal{G}_T|^2/\Delta$, characteristic of the transmon, is accompanied by an additional higher-order shift $\chi_M=2E_M |\mathcal{G}_o|^2/\Delta^2$, which dispersively couples the photon, transmon, and parity degrees of freedom. Focusing on the $k=0$ sector, we obtain the Hamiltonian for the qubit with the parity-dependent dispersive interaction
$H_{\text{eff},q} = E_M\tau_z + a^\dag a \left(\omega_c + \chi_T-\chi_M\tau_z\right).$
We realize that the cavity resonance frequency of $0\rightarrow 1$ photon transition is dependent on the parity of the ground state. This interaction, which scales as $\chi_M$, can be used in a homodyne measurement setup to determine which of the two ground states the MT occupies \cite{PhysRevA.76.012325,PhysRevLett.94.123602,Johnson2010,Fragner28112008,PhysRevLett.95.060501}, opening the way to use this pair of states as a qubit. The signal-to-noise ratio of such a readout scheme depends also on the strength of the decoherence processes, the scale of $E_M$, and the quality factor of the cavity.

\subsection{Qubit initialization and control}
In the presence of a cavity-MT interaction, we denote by $|\overline{1,s;0}\rangle$ and $|\overline{0,s;1}\rangle$ the dressed states of the $\ell=1$ excitation sector, which are approximate  eigenstates of Eq.~(\ref{eq:jc-N photons block}) in the dispersive regime. We drive the qubit via the cavity with $\xi(t)(a+a^\dagger)$, written in terms of the itinerant electric field at the port of the resonator with the drive amplitude $\xi(t)$. The drive mixes the dressed states $|\overline{1,s;0}\rangle$ with the bare states $|0,s';0 \rangle$ according to
\begin{equation}
\label{eq:drive-indirect}
\begin{aligned}
H_D^{\text{cav}} &= \xi(t)\sum_s\left(\frac{\mathcal{G}_o}{\Delta}|0,s;0\rangle \langle \overline{1,s;0}|\right.
\\
&\left.+s\frac{\mathcal{G}_x}{\Delta}|0,s;0\rangle \langle \overline{1,-s;0}| \right) +\text{h.c.},
\end{aligned}
\end{equation}
To first order in $|\mathcal{G}_{o,x}|$,  this scaling does not depend on $E_M$ and is similar to the transmon-cavity case. The general dependence on the small parameters $|\mathcal{G}_{o,x}|/\Delta$ and $E_M/\Delta$ is different, with higher-order terms depending on $E_M$. Therefore, in the first order, this form of the drive would evidently lead to a qualitatively similar spectroscopic pattern of the transition strengths on $n_g$.

In addition to the logical qubit $\ell=0$, $s=\pm$ (which lacks direct dipole couplings between its two states), we denote the lowest dressed doublet $|\overline{1,s;0}\rangle$ of the $\ell=1$ sector as the control doublet [see Fig.~\ref*{fig:cooling-control}(c)], forming together a double $\Lambda$ system (with shared ground states). Due to the small energy splitting between the logical qubit's levels, the equilibrium thermal state of the system generically mixes the two levels. To prepare the qubit in a pure state, a cooling procedure should initially be performed. This involves an external drive operating at the frequency $\omega_{++}$, where $\omega_{ss'}$ corresponds to $|\overline{1,s;0}\rangle\rightleftarrows |0,s';0\rangle$ transitions. Photonic decay channels exist between the control states and the qubit states, with the decay rates coinciding with the dipole transitions. For transitions that preserve $s$, the decay rates are $\gamma_o = \gamma |\cos(\eta_1-\eta_0)|^2$ (neglecting the small energy difference between the two transitions), and for transitions that flip $s$, we take $\gamma_x = \gamma |\sin(\eta_1-\eta_0)|^2$, with $\gamma$ as the decay rate associated with the transmon. The process results in the relaxation of the system to the ground state, as depicted in Fig.~\ref{fig:cooling-control}(a). 

Following the initialization to the pure state $|0,-;0\rangle$, single-qubit quantum gates can be performed on the qubit. To demonstrate a simple gate operation, we focus here on a population flip, taking $|0,-;0\rangle\to|0,+;0\rangle$. The trick is to use a two-tone microwave photon drive operating in the frequencies $\omega_{+-}-\delta$ and $\omega_{--}-\delta$ sharing the same detuning $\delta$, fixed between the control doublet levels. This results in a coherent population transfer between the states of the qubit; see Fig.~\ref{fig:cooling-control}(b). A complete transition is achieved by taking $\delta \simeq E_M$, in which case the probability for the occupation of $|0,+;0\rangle$ takes the form $P(t)\simeq \sin^2\left(\frac{\xi^2|\mathcal{G}_o\mathcal{G}_x|}{2E_M\Delta^2}t\right)$ \cite{cohen1998atom}.
The transition is a coherent evolution where a dynamical phase is accumulated, which can be described as a combination of $R_x$ and $R_z$ rotations.

\section{Discussion}
The MT may show a remarkable resilience to major forms of decoherence which affect the transmon. It can be protected from photon-induced dephasing by tuning to the $n_g\to 1/4$ point following each qubit operation. At this point, the effective coupling to the cavity is turned off as $\mathcal{G}_0\to 0$, leading to the vanishing of the qubit-photon interaction term in the dispersive Hamiltonian. A similar effect is achieved by increasing the ratio $E_J/E_C$, as seen in Fig.~\ref{fig:transmon-setup}.
In addition, by operating in the ground-state sector of the transmon, with no discernible direct dipole coupling, the MT is not affected by spontaneous emission or transmon relaxation processes. Future research is needed into the influence of nonequilibrium quasi-particles and topological protection. Finally, recent experimental progress indicates that the global parity has a very long lifetime, exceeding $10$ms \cite{2015arXiv150105155H}. This process will be setting the upper bound for the MT coherence time.

\begin{acknowledgements}
K.Y. and E.Gr. acknowledge the support from the Israel Science Foundation (Grant No. 401/12) and the European Union’s Seventh Framework Programme (FP7/2007-2013) under Grant No. 303742. E.Gi. acknowledges support from EPSRC (Grant No. EP/I026231/1). E.Gr. and E.Gi. acknowledge support from the Royal Society International Exchanges programme, Grant No. IE121282. Details of the data and how to request access are available from the University of Surrey publications repository doi: 10.15126/surreydata.00808295.
\end{acknowledgements}

\appendix*
\setcounter{equation}{0}
\section*{APPENDIX}
Here we provide an asymptotic solution, based on the WKB method \cite{1929goldstein,1976JPhB....9L.513C,1923jeffreys}, to the equation $H_T f(\vphi)=E f(\vphi)$ with the boundary condition $f(\vphi+2\pi) = e^{i\theta}f(\vphi)$, ${(\theta \in \mathbb{R})}$, which we write as
\begin{equation}	
\label{eq:sc-equation 1st form}
	f''(\vphi) +\left(\frac{E}{4E_C}+\frac{E_J}{4E_C}\cos(\vphi)\right)f(\vphi) = 0.
\end{equation}
We are interested in the transmon regime $E_J/E_C\gg 1$ where the fluctuations of $\vphi$ are mostly localized around $\vphi=0$ and the energy has a small deviation $\delta E_k$ from the harmonic-oscillator values: $E_k = -E_J+\sqrt{8E_C E_J}\left(k+1/2\right)+\delta E_k$, where $k=0,1,2,\ldots$ and $\delta E_k\ll \sqrt{8 E_C E_J}$. By inserting this expression into Eq.~(\ref{eq:sc-equation 1st form}) and rearranging, we get
\begin{equation}
\label{eq:sc-equation 2nd form}
	f''(\vphi) +\sqrt{\frac{E_J}{2E_C}}\left(\nu + \frac{1}{2} -\sqrt{\frac{E_J}{2E_C}}\sin^2(\vphi/2)\right) f(\vphi) = 0,
\end{equation}
where $\nu = k + \delta E_k / \sqrt{8E_C E_J}$. As $E_J$ increases, $\nu$ approaches an integer value.

In deriving the solution, we focus for convenience on the domain $-\pi<\vphi<\pi$, which contains two barriers centered at $\vphi = \pm \pi$. To describe tunneling through the barriers, the wave function should be a linear combination of two independent functions: one exponentially increasing and one exponentially decaying. For these, we assume the form $\phi_{\pm}(\vphi) = A_\pm (\vphi) e^{\pm S(\vphi)}$, where $S(\vphi)$ is the action through the barrier, for which we take the ansatz $S(\vphi) = \sqrt{\frac{2E_J}{E_C}}\cos(\vphi/2)$. By inserting $\phi_\pm(\vphi)$ into Eq.~(\ref{eq:sc-equation 2nd form}) and neglecting terms of order $\sim\sqrt{E_C/E_J}$, we obtain a first-order equation
\begin{equation}
\label{eq:sc-equation for the amplitude}
	A_\pm(\vphi)\left(\cos(\vphi/2) \mp (2\nu+1)\right) + 4A_\pm'(\vphi) \sin(\vphi/2)=0,
\end{equation}
which is readily solved to find the two independent solutions
\begin{equation}
\label{eq:sc-two independent solutions}
	\phi_\pm (\vphi) = \frac{\tan(\vphi/4)^{\pm(\nu+1/2)}}{\sqrt{\sin(\vphi/2)}} e^{\pm \sqrt{2E_J/E_C} \cos(\vphi/2)}.
\end{equation}
These are valid mainly close to $\vphi = \pm \pi$. Close to $\vphi = 0$, we can rewrite Eq.~(\ref{eq:sc-equation 2nd form}) approximately as
\begin{equation}
\label{eq:sc-equation approximation near phi = 0}
	\frac{d^2f(z)}{dz^2}+ \left(\nu+\frac{1}{2}-\frac{z^2}{4}\right)f(z) = 0,
\end{equation}
where $z = (E_J/2E_C)^{1/4}\vphi$. Equation~(\ref{eq:sc-equation approximation near phi = 0}) is the Weber equation, which has two independent solutions $D_\nu(z)$ and $D_{-(\nu+1)}(iz)$, i.e., the parabolic cylinder functions (following the notation of Abramowitz and Stegun \cite{abramowitz}). For positive integer $\nu$, this equation simply describes the harmonic oscillator.

We can completely satisfy the solution in the domain $-\pi < \vphi < \pi$ using the three functions
\begin{equation}
\label{eq:sc-form of solution}
\begin{aligned}
	&f_L(\vphi) = A_L \phi_+(\vphi) + B_L \phi_-(\vphi), \\
	&f_M(\vphi) = A_M \Psi(\vphi) + B_M \Omega(\vphi), \\
	&f_R(\vphi) = A_R \phi_+(\vphi) + B_R \phi_-(\vphi), \\
\end{aligned}
\end{equation}
where $\Psi(\vphi) = D_\nu(z)$ and $\Omega(\vphi) = D_{-(\nu+1)}(iz)$ (for simplicity we omit the band index $\nu$ from the basis functions). Here, $f_L$ ($f_R$) is the solution to the left (right) of $\vphi = 0$ and $f_M$ is valid in the  region close to $\vphi = 0$. In order to impose the boundary condition on Eq.~(\ref{eq:sc-form of solution}), we need to represent the coefficients of $f_R$ as a linear combination of the coefficients of $f_L$. This is achieved by comparing $\phi_\pm(\vphi)$ to $\Psi(\vphi)$ and $\Omega(\vphi)$ in their common region of validity. 

Using an asymptotic approximation for the parabolic cylinder functions \cite{abramowitz}, for any $\nu$ and $z\gg 1$ we obtain the form
\begin{equation}
\label{eq:sc-approximation of parabolic functions}
\begin{aligned}
	&\Psi(\vphi) \simeq \left(\frac{E_J}{2E_C}\right)^{\frac{\nu}{4}}\vphi^\nu e^{-\sqrt{\frac{E_J}{32E_C}}\vphi^2},\\
	&\Omega(\vphi) \simeq e^{-i\frac{\pi}{2}(\nu+1)}\left(\frac{E_J}{2E_C}\right)^{-\frac{(\nu+1)}{4}}\vphi^{-(\nu+1)} e^{\sqrt{\frac{E_J}{32E_C}}\vphi^2}.
\end{aligned}
\end{equation}
In addition, Eq.~(\ref{eq:sc-two independent solutions}) takes the approximate form near $\vphi=0$, 
\begin{equation}
\label{eq:sc-approximation for phi basis}
\begin{aligned}
	&\phi_+(\vphi) \simeq 2^{-\left(2\nu+\frac{1}{2}\right)}e^{\sqrt{\frac{2E_J}{E_C}}}\vphi^\nu e^{-\sqrt{\frac{E_J}{32E_C}}\vphi^2}, \\
	&\phi_-(\vphi) \simeq 2^{\left(2\nu+\frac{3}{2}\right)}e^{-\sqrt{\frac{2E_J}{E_C}}}\vphi^{-\nu}|\vphi|^{-1} e^{\sqrt{\frac{E_J}{32E_C}}\vphi^2}.
\end{aligned}
\end{equation}
Writing Eq.~(\ref{eq:sc-approximation for phi basis}) in terms of Eq.~(\ref{eq:sc-approximation of parabolic functions}) for $\vphi \gtrsim 0$, we get
\begin{equation}
\label{eq:sc-representation of phi > 0 using parabolic}
\begin{aligned}
	&\phi_+(\vphi) = 2^{-(2\nu+\frac{1}{2})} \left(\frac{E_J}{2E_C}\right)^{-\frac{\nu}{4}}
	e^{\sqrt{\frac{2E_J}{E_C}}}\Psi(\vphi), \\
	&\phi_-(\vphi) =2^{(2\nu+\frac{3}{2})} e^{i\frac{\pi}{2}(\nu+1)}\left(\frac{E_J}{2E_C}\right)^{\frac{\nu+1}{4}}
	e^{-\sqrt{\frac{2E_J}{E_C}}}\Omega(\vphi). \\
\end{aligned}
\end{equation}
Using the above, $f_R(\vphi)$ can be expressed approximately using $\Psi(\vphi)$, $\Omega(\vphi)$ for $\varphi\gtrsim 0$.
A similar method can be applied in the region $\vphi \lesssim 0$, but the approximations in Eq.~(\ref{eq:sc-approximation of parabolic functions}) cannot be used directly since they are valid only for $\vphi>0$. Instead we use the identities \citep{abramowitz}
\begin{equation}
\label{eq:sc-parabolic identities for negative argument}
\begin{aligned}
	&\Psi(-\vphi) = e^{i\pi \nu}\Psi(\vphi) - \frac{\sqrt{2\pi}}{\Gamma(-\nu)}e^{i\frac{\pi}{2}(\nu-1)}\Omega(\vphi),\\
	&\Omega(-\vphi) = e^{i\pi (\nu+1)}\Omega(\vphi) + \frac{\sqrt{2\pi}}{\Gamma(\nu+1)}e^{i\frac{\pi}{2}\nu}\Psi(\vphi).\\
\end{aligned}
\end{equation}
Writing $\phi_\pm(\vphi)$ in the region $\vphi \lesssim 0$ as $\phi_\pm(-|\vphi|)$, we get
\begin{equation}
\label{eq:sc-representation of phi < 0 using parabolic}
\begin{aligned}
	&\phi_+(\vphi) = e^{i\pi \nu} 2^{-(2\nu+\frac{1}{2})} \left(\frac{E_J}{2E_C}\right)^{-\frac{\nu}{4}}
	e^{\sqrt{\frac{2E_J}{E_C}}}\Psi(-\vphi), \\
	&\phi_-(\vphi) = e^{-i\frac{\pi}{2}(\nu-1)}2^{(2\nu+\frac{3}{2})} \left(\frac{E_J}{2E_C}\right)^{\frac{\nu+1}{4}}
	e^{-\sqrt{\frac{2E_J}{E_C}}}\Omega(-\vphi). \\
\end{aligned}
\end{equation}
Using Eq.~(\ref{eq:sc-parabolic identities for negative argument}), we can now also represent $f_L(\vphi)$ using $\Psi(\vphi)$, $\Omega(\vphi)$. By comparing the coefficients of this basis in $f_R(\vphi)$ to the coefficients in $f_L(\vphi)$, we acquire the connection matrix
\begin{equation}
\label{eq:sc-connection matrix}	
	\begin{pmatrix}
		A_R \\ B_R
	\end{pmatrix}
	=
	\begin{pmatrix}
		e^{2\pi i \nu} & 0 \\
		\zeta e^{i\pi \nu} & e^{i\pi}
	\end{pmatrix}
	\begin{pmatrix}
		A_L \\ B_L
	\end{pmatrix}
	\equiv
	\mathcal{C}
	\begin{pmatrix}
		A_L \\ B_L
	\end{pmatrix},
\end{equation}
where we neglected the wave-function terms containing $e^{-\sqrt{8E_J/E_C}}$. The factor $\zeta$ is
\begin{equation}
\label{eq:sc-zeta}
\zeta=2^{-(4\nu+2)}\left(\frac{E_J}{2E_C}\right)^{-\frac{\nu}{2}-\frac{1}{4}}e^{\sqrt{\frac{8E_J}{E_C}}} 
\frac{\sqrt{2\pi}}{\Gamma(-\nu)}.
\end{equation}
Next we construct a matrix which represents the boundary condition via the coefficients by using the property of Eq.~(\ref{eq:sc-two independent solutions}): $\phi_{\pm}(\vphi+2\pi) = e^{\pm i\pi\nu}\phi_{\mp}(\vphi)$. Together with the required boundary condition, we obtain
\begin{equation}
\label{eq:sc-boundary condition matrix}
\begin{pmatrix}
		A_R \\ B_R
	\end{pmatrix}
	=
	\begin{pmatrix}
		0 & e^{-i(\pi\nu-\theta)} \\
		 e^{i(\pi\nu+\theta)} & 0
	\end{pmatrix}
	\begin{pmatrix}
		A_L \\ B_L
	\end{pmatrix}
	\equiv
	\mathcal{B}
	\begin{pmatrix}
		A_L \\ B_L
	\end{pmatrix}.
\end{equation}
Combining Eq.~(\ref{eq:sc-connection matrix}) with Eq.~(\ref{eq:sc-boundary condition matrix}) gives us a system of equations for the coefficients, which has a nontrivial solution only when det$\left(\mathcal{C}-\mathcal{B}\right)=0$. Using the approximation $\nu\simeq k$ [everywhere except in $\Gamma(-\nu)$, which we deal with separately below], the condition can be written as $\zeta = 2\cos(\theta)$. In order to retrieve the value of $\delta E_k$, we next use the identity 
$\Gamma(-\nu)\Gamma(\nu+1) = -\frac{\pi}{\sin(\pi \nu)}$ \cite{abramowitz} to get, expanding to first order in $\delta E_k$ in the denominator,
\begin{equation}
\label{eq:sc-delta using Gamma}
	\Gamma(-\nu) \simeq (-1)^{k+1} \frac{\sqrt{8E_CE_J}}{\Gamma(\nu+1)\delta E_k}. 
\end{equation}
By plugging this expression into Eq.~(\ref{eq:sc-zeta}) we obtain the tight-binding-like spectrum $\delta E_k=t_k\cos(\theta)$ with the ``tunneling amplitude'' $t_k$ defined as
\begin{equation}
\label{eq:sc-tunnelling term}
	t_k=(-1)^{k+1}\frac{2^{4(k+1)}E_C}{k!} \sqrt{\frac{2}{\pi}}\left(\frac{E_J}{2E_C}\right)^{\frac{k}{2}+\frac{3}{4}}e^{-\sqrt{\frac{8E_J}{E_C}}}.
\end{equation}

The functions described in Eq.~(\ref{eq:sc-form of solution}) outline the entire solution in the region $-\pi <\vphi < \pi$. Although we only found the coefficients of $f_L(\vphi)$ and $f_R(\vphi)$, a similar process can be applied to find the connections to $f_M(\vphi)$. In keeping with the spirit of our approximation, we argue that in cases where the tunneling process between the barriers is negligible, $f_M(\vphi)$, which need not satisfy the relevant boundary condition, can be used as the wave function of the transmon to a good approximation. Finally, since $\Omega(\vphi)$ is a bounded function in the region $-\pi<\vphi<\pi$ and the ratio $B_M/A_M \sim e^{-\sqrt{8E_J/E_C}+i\theta}$ is exponentially small, we can approximate $f_M(\vphi) \simeq \Psi_k(\vphi)$, where $\Psi_k(\vphi)$ is the $k$th harmonic-oscillator wave function. Since the dependence of $f_M$ on $\theta$ appears only as a relative phase between its two coefficients, in this approximation the offset charge is absent from the wave function and the domain of integration can be taken as $-\infty < \vphi < \infty$. Using this wave function, we form the basis for the rest of our analysis.
%%%%%%%%%%%%%%%%%%%%%%%%%%%%%%%%%%%%%%%%%%%%%%%%%%%%%%%%%%%%%%%%%%%%%%%%%%%%%%%%%%%%%%%%%%%%
%

\begin{thebibliography}{41}%
\makeatletter
\providecommand \@ifxundefined [1]{%
 \@ifx{#1\undefined}
}%
\providecommand \@ifnum [1]{%
 \ifnum #1\expandafter \@firstoftwo
 \else \expandafter \@secondoftwo
 \fi
}%
\providecommand \@ifx [1]{%
 \ifx #1\expandafter \@firstoftwo
 \else \expandafter \@secondoftwo
 \fi
}%
\providecommand \natexlab [1]{#1}%
\providecommand \enquote  [1]{``#1''}%
\providecommand \bibnamefont  [1]{#1}%
\providecommand \bibfnamefont [1]{#1}%
\providecommand \citenamefont [1]{#1}%
\providecommand \href@noop [0]{\@secondoftwo}%
\providecommand \href [0]{\begingroup \@sanitize@url \@href}%
\providecommand \@href[1]{\@@startlink{#1}\@@href}%
\providecommand \@@href[1]{\endgroup#1\@@endlink}%
\providecommand \@sanitize@url [0]{\catcode `\\12\catcode `\$12\catcode
  `\&12\catcode `\#12\catcode `\^12\catcode `\_12\catcode `\%12\relax}%
\providecommand \@@startlink[1]{}%
\providecommand \@@endlink[0]{}%
\providecommand \url  [0]{\begingroup\@sanitize@url \@url }%
\providecommand \@url [1]{\endgroup\@href {#1}{\urlprefix }}%
\providecommand \urlprefix  [0]{URL }%
\providecommand \Eprint [0]{\href }%
\providecommand \doibase [0]{http://dx.doi.org/}%
\providecommand \selectlanguage [0]{\@gobble}%
\providecommand \bibinfo  [0]{\@secondoftwo}%
\providecommand \bibfield  [0]{\@secondoftwo}%
\providecommand \translation [1]{[#1]}%
\providecommand \BibitemOpen [0]{}%
\providecommand \bibitemStop [0]{}%
\providecommand \bibitemNoStop [0]{.\EOS\space}%
\providecommand \EOS [0]{\spacefactor3000\relax}%
\providecommand \BibitemShut  [1]{\csname bibitem#1\endcsname}%
\let\auto@bib@innerbib\@empty
%</preamble>
\bibitem [{\citenamefont {Koch}\ \emph {et~al.}(2007)\citenamefont {Koch},
  \citenamefont {Yu}, \citenamefont {Gambetta}, \citenamefont {Houck},
  \citenamefont {Schuster}, \citenamefont {Majer}, \citenamefont {Blais},
  \citenamefont {Devoret}, \citenamefont {Girvin},\ and\ \citenamefont
  {Schoelkopf}}]{koch_charge-insensitive_2007}%
  \BibitemOpen
  \bibfield  {author} {\bibinfo {author} {\bibfnamefont {J.}~\bibnamefont
  {Koch}}, \bibinfo {author} {\bibfnamefont {T.~M.}\ \bibnamefont {Yu}},
  \bibinfo {author} {\bibfnamefont {J.}~\bibnamefont {Gambetta}}, \bibinfo
  {author} {\bibfnamefont {A.~A.}\ \bibnamefont {Houck}}, \bibinfo {author}
  {\bibfnamefont {D.~I.}\ \bibnamefont {Schuster}}, \bibinfo {author}
  {\bibfnamefont {J.}~\bibnamefont {Majer}}, \bibinfo {author} {\bibfnamefont
  {A.}~\bibnamefont {Blais}}, \bibinfo {author} {\bibfnamefont {M.~H.}\
  \bibnamefont {Devoret}}, \bibinfo {author} {\bibfnamefont {S.~M.}\
  \bibnamefont {Girvin}}, \ and\ \bibinfo {author} {\bibfnamefont {R.~J.}\
  \bibnamefont {Schoelkopf}},\ }\href {\doibase 10.1103/PhysRevA.76.042319}
  {\bibfield  {journal} {\bibinfo  {journal} {Phys. Rev. A}\ }\textbf {\bibinfo
  {volume} {76}},\ \bibinfo {pages} {042319} (\bibinfo {year}
  {2007})}\BibitemShut {NoStop}%
\bibitem [{\citenamefont {Schreier}\ \emph {et~al.}(2008)\citenamefont
  {Schreier}, \citenamefont {Houck}, \citenamefont {Koch}, \citenamefont
  {Schuster}, \citenamefont {Johnson}, \citenamefont {Chow}, \citenamefont
  {Gambetta}, \citenamefont {Majer}, \citenamefont {Frunzio}, \citenamefont
  {Devoret}, \citenamefont {Girvin},\ and\ \citenamefont
  {Schoelkopf}}]{schreier2008suppressing}%
  \BibitemOpen
  \bibfield  {author} {\bibinfo {author} {\bibfnamefont {J.~A.}\ \bibnamefont
  {Schreier}}, \bibinfo {author} {\bibfnamefont {A.~A.}\ \bibnamefont {Houck}},
  \bibinfo {author} {\bibfnamefont {J.}~\bibnamefont {Koch}}, \bibinfo {author}
  {\bibfnamefont {D.~I.}\ \bibnamefont {Schuster}}, \bibinfo {author}
  {\bibfnamefont {B.~R.}\ \bibnamefont {Johnson}}, \bibinfo {author}
  {\bibfnamefont {J.~M.}\ \bibnamefont {Chow}}, \bibinfo {author}
  {\bibfnamefont {J.~M.}\ \bibnamefont {Gambetta}}, \bibinfo {author}
  {\bibfnamefont {J.}~\bibnamefont {Majer}}, \bibinfo {author} {\bibfnamefont
  {L.}~\bibnamefont {Frunzio}}, \bibinfo {author} {\bibfnamefont {M.~H.}\
  \bibnamefont {Devoret}}, \bibinfo {author} {\bibfnamefont {S.~M.}\
  \bibnamefont {Girvin}}, \ and\ \bibinfo {author} {\bibfnamefont {R.~J.}\
  \bibnamefont {Schoelkopf}},\ }\href {\doibase 10.1103/PhysRevB.77.180502}
  {\bibfield  {journal} {\bibinfo  {journal} {Phys. Rev. B}\ }\textbf {\bibinfo
  {volume} {77}},\ \bibinfo {pages} {180502} (\bibinfo {year}
  {2008})}\BibitemShut {NoStop}%
\bibitem [{\citenamefont {Houck}\ \emph {et~al.}(2008)\citenamefont {Houck},
  \citenamefont {Schreier}, \citenamefont {Johnson}, \citenamefont {Chow},
  \citenamefont {Koch}, \citenamefont {Gambetta}, \citenamefont {Schuster},
  \citenamefont {Frunzio}, \citenamefont {Devoret}, \citenamefont {Girvin}
  \emph {et~al.}}]{houck2008controlling}%
  \BibitemOpen
  \bibfield  {author} {\bibinfo {author} {\bibfnamefont {A.~A.}\ \bibnamefont
  {Houck}}, \bibinfo {author} {\bibfnamefont {J.~A.}\ \bibnamefont {Schreier}},
  \bibinfo {author} {\bibfnamefont {B.~R.}\ \bibnamefont {Johnson}}, \bibinfo
  {author} {\bibfnamefont {J.~M.}\ \bibnamefont {Chow}}, \bibinfo {author}
  {\bibfnamefont {J.}~\bibnamefont {Koch}}, \bibinfo {author} {\bibfnamefont
  {J.}~\bibnamefont {Gambetta}}, \bibinfo {author} {\bibfnamefont {D.~I.}\
  \bibnamefont {Schuster}}, \bibinfo {author} {\bibfnamefont {L.}~\bibnamefont
  {Frunzio}}, \bibinfo {author} {\bibfnamefont {M.~H.}\ \bibnamefont
  {Devoret}}, \bibinfo {author} {\bibfnamefont {S.~M.}\ \bibnamefont {Girvin}},
   \emph {et~al.},\ }\href@noop {} {\bibfield  {journal} {\bibinfo  {journal}
  {Phys. Rev. Lett.}\ }\textbf {\bibinfo {volume} {101}},\ \bibinfo {pages}
  {080502} (\bibinfo {year} {2008})}\BibitemShut {NoStop}%
\bibitem [{\citenamefont {Read}\ and\ \citenamefont
  {Green}(2000)}]{read2000paired}%
  \BibitemOpen
  \bibfield  {author} {\bibinfo {author} {\bibfnamefont {N.}~\bibnamefont
  {Read}}\ and\ \bibinfo {author} {\bibfnamefont {D.}~\bibnamefont {Green}},\
  }\href@noop {} {\bibfield  {journal} {\bibinfo  {journal} {Phys. Rev.
  B}\ }\textbf {\bibinfo {volume} {61}},\ \bibinfo {pages} {10267} (\bibinfo
  {year} {2000})}\BibitemShut {NoStop}%
\bibitem [{\citenamefont {Kitaev}(2001)}]{kitaev2007unpaired}%
  \BibitemOpen
  \bibfield  {author} {\bibinfo {author} {\bibfnamefont {A.~Y.}\ \bibnamefont
  {Kitaev}},\ }\href@noop {} {\bibfield  {journal} {\bibinfo  {journal}
  {Phys. Usp.}\ }\textbf {\bibinfo {volume} {44}},\ \bibinfo {pages} {131}
  (\bibinfo {year} {2007})}\BibitemShut {NoStop}%
\bibitem [{\citenamefont {Hassler}\ \emph {et~al.}(2011)\citenamefont
  {Hassler}, \citenamefont {Akhmerov},\ and\ \citenamefont
  {Beenakker}}]{hassler2011top}%
  \BibitemOpen
  \bibfield  {author} {\bibinfo {author} {\bibfnamefont {F.}~\bibnamefont
  {Hassler}}, \bibinfo {author} {\bibfnamefont {A.~R.}\ \bibnamefont
  {Akhmerov}}, \ and\ \bibinfo {author} {\bibfnamefont {C.~W.~J.}\ \bibnamefont
  {Beenakker}},\ }\href@noop {} {\bibfield  {journal} {\bibinfo  {journal} {New
  J. Phys.}\ }\textbf {\bibinfo {volume} {13}},\ \bibinfo {pages}
  {095004} (\bibinfo {year} {2011})}\BibitemShut {NoStop}%
\bibitem [{\citenamefont {Bonderson}\ and\ \citenamefont
  {Lutchyn}(2011)}]{bonderson2011topological}%
  \BibitemOpen
  \bibfield  {author} {\bibinfo {author} {\bibfnamefont {P.}~\bibnamefont
  {Bonderson}}\ and\ \bibinfo {author} {\bibfnamefont {R.~M.}\ \bibnamefont
  {Lutchyn}},\ }\href@noop {} {\bibfield  {journal} {\bibinfo  {journal} {Phys.
  Rev. Lett.}\ }\textbf {\bibinfo {volume} {106}},\ \bibinfo {pages} {130505}
  (\bibinfo {year} {2011})}\BibitemShut {NoStop}%
\bibitem [{\citenamefont {van Heck}\ \emph
  {et~al.}(2011{\natexlab{a}})\citenamefont {van Heck}, \citenamefont
  {Hassler}, \citenamefont {Akhmerov},\ and\ \citenamefont
  {Beenakker}}]{van2011coulomb}%
  \BibitemOpen
  \bibfield  {author} {\bibinfo {author} {\bibfnamefont {B.}~\bibnamefont {van
  Heck}}, \bibinfo {author} {\bibfnamefont {F.}~\bibnamefont {Hassler}},
  \bibinfo {author} {\bibfnamefont {A.~R.}\ \bibnamefont {Akhmerov}}, \ and\
  \bibinfo {author} {\bibfnamefont {C.~W.~J.}\ \bibnamefont {Beenakker}},\
  }\href@noop {} {\bibfield  {journal} {\bibinfo  {journal} {Phys. Rev. B}\
  }\textbf {\bibinfo {volume} {84}},\ \bibinfo {pages} {180502} (\bibinfo
  {year} {2011}{\natexlab{a}})}\BibitemShut {NoStop}%
\bibitem [{\citenamefont {{Schmidt}}\ \emph {et~al.}(2013)\citenamefont
  {{Schmidt}}, \citenamefont {{Nunnenkamp}},\ and\ \citenamefont
  {{Bruder}}}]{schmidt2013majorana}%
  \BibitemOpen
  \bibfield  {author} {\bibinfo {author} {\bibfnamefont {T.~L.}\ \bibnamefont
  {{Schmidt}}}, \bibinfo {author} {\bibfnamefont {A.}~\bibnamefont
  {{Nunnenkamp}}}, \ and\ \bibinfo {author} {\bibfnamefont {C.}~\bibnamefont
  {{Bruder}}},\ }\href {\doibase 10.1103/PhysRevLett.110.107006} {\bibfield
  {journal} {\bibinfo  {journal} {Phys. Rev. Lett.}\ }\textbf {\bibinfo
  {volume} {110}},\ \bibinfo {eid} {107006} (\bibinfo {year}
  {2013})}\BibitemShut {NoStop}%
\bibitem [{\citenamefont {Pekker}\ \emph {et~al.}(2013)\citenamefont {Pekker},
  \citenamefont {Hou}, \citenamefont {Manucharyan},\ and\ \citenamefont
  {Demler}}]{pekker2013proposal}%
  \BibitemOpen
  \bibfield  {author} {\bibinfo {author} {\bibfnamefont {D.}~\bibnamefont
  {Pekker}}, \bibinfo {author} {\bibfnamefont {C.-Y.}\ \bibnamefont {Hou}},
  \bibinfo {author} {\bibfnamefont {V.~E.}\ \bibnamefont {Manucharyan}}, \ and\
  \bibinfo {author} {\bibfnamefont {E.}~\bibnamefont {Demler}},\ }\href
  {\doibase 10.1103/PhysRevLett.111.107007} {\bibfield  {journal} {\bibinfo
  {journal} {Phys. Rev. Lett.}\ }\textbf {\bibinfo {volume} {111}},\ \bibinfo
  {pages} {107007} (\bibinfo {year} {2013})}\BibitemShut {NoStop}%
\bibitem [{\citenamefont {M\"uller}\ \emph {et~al.}(2013)\citenamefont
  {M\"uller}, \citenamefont {Bourassa},\ and\ \citenamefont
  {Blais}}]{blais_majorana_2013}%
  \BibitemOpen
  \bibfield  {author} {\bibinfo {author} {\bibfnamefont {C.}~\bibnamefont
  {M\"uller}}, \bibinfo {author} {\bibfnamefont {J.}~\bibnamefont {Bourassa}},
  \ and\ \bibinfo {author} {\bibfnamefont {A.}~\bibnamefont {Blais}},\ }\href
  {\doibase 10.1103/PhysRevB.88.235401} {\bibfield  {journal} {\bibinfo
  {journal} {Phys. Rev. B}\ }\textbf {\bibinfo {volume} {88}},\ \bibinfo
  {pages} {235401} (\bibinfo {year} {2013})}\BibitemShut {NoStop}%
\bibitem [{\citenamefont {Cottet}\ \emph {et~al.}(2013)\citenamefont {Cottet},
  \citenamefont {Kontos},\ and\ \citenamefont
  {Dou\c{c}ot}}]{cottet_majorana_squeezing_2013}%
  \BibitemOpen
  \bibfield  {author} {\bibinfo {author} {\bibfnamefont {A.}~\bibnamefont
  {Cottet}}, \bibinfo {author} {\bibfnamefont {T.}~\bibnamefont {Kontos}}, \
  and\ \bibinfo {author} {\bibfnamefont {B.}~\bibnamefont {Dou\c{c}ot}},\
  }\href@noop {} {\bibfield  {journal} {\bibinfo  {journal} {Phys. Rev B}\
  }\textbf {\bibinfo {volume} {88}},\ \bibinfo {pages} {195415} (\bibinfo
  {year} {2013})}\BibitemShut {NoStop}%
\bibitem [{\citenamefont {{Ginossar}}\ and\ \citenamefont
  {{Grosfeld}}(2014)}]{GinossarGrosfeld2014}%
  \BibitemOpen
  \bibfield  {author} {\bibinfo {author} {\bibfnamefont {E.}~\bibnamefont
  {{Ginossar}}}\ and\ \bibinfo {author} {\bibfnamefont {E.}~\bibnamefont
  {{Grosfeld}}},\ }\href@noop {} {\bibfield  {journal} {\bibinfo  {journal}
  {Nat. Commun.}\ }\textbf {\bibinfo {volume} {5}},\ \bibinfo {pages} {4772}
  (\bibinfo {year} {2014})}\BibitemShut {NoStop}%
\bibitem [{\citenamefont {Ohm}\ and\ \citenamefont
  {Hassler}(2015)}]{PhysRevB.91.085406}%
  \BibitemOpen
  \bibfield  {author} {\bibinfo {author} {\bibfnamefont {C.}~\bibnamefont
  {Ohm}}\ and\ \bibinfo {author} {\bibfnamefont {F.}~\bibnamefont {Hassler}},\
  }\href {\doibase 10.1103/PhysRevB.91.085406} {\bibfield  {journal} {\bibinfo
  {journal} {Phys. Rev. B}\ }\textbf {\bibinfo {volume} {91}},\ \bibinfo
  {pages} {085406} (\bibinfo {year} {2015})}\BibitemShut {NoStop}%
\bibitem [{\citenamefont {Blais}\ \emph {et~al.}(2004)\citenamefont {Blais},
  \citenamefont {Huang}, \citenamefont {Wallraff}, \citenamefont {Girvin},\
  and\ \citenamefont {Schoelkopf}}]{PhysRevA.69.062320}%
  \BibitemOpen
  \bibfield  {author} {\bibinfo {author} {\bibfnamefont {A.}~\bibnamefont
  {Blais}}, \bibinfo {author} {\bibfnamefont {R.-S.}\ \bibnamefont {Huang}},
  \bibinfo {author} {\bibfnamefont {A.}~\bibnamefont {Wallraff}}, \bibinfo
  {author} {\bibfnamefont {S.~M.}\ \bibnamefont {Girvin}}, \ and\ \bibinfo
  {author} {\bibfnamefont {R.~J.}\ \bibnamefont {Schoelkopf}},\ }\href
  {\doibase 10.1103/PhysRevA.69.062320} {\bibfield  {journal} {\bibinfo
  {journal} {Phys. Rev. A}\ }\textbf {\bibinfo {volume} {69}},\ \bibinfo
  {pages} {062320} (\bibinfo {year} {2004})}\BibitemShut {NoStop}%
\bibitem [{\citenamefont {Gambetta}\ \emph {et~al.}(2006)\citenamefont
  {Gambetta}, \citenamefont {Blais}, \citenamefont {Schuster}, \citenamefont
  {Wallraff}, \citenamefont {Frunzio}, \citenamefont {Majer}, \citenamefont
  {Devoret}, \citenamefont {Girvin},\ and\ \citenamefont
  {Schoelkopf}}]{PhysRevA.74.042318}%
  \BibitemOpen
  \bibfield  {author} {\bibinfo {author} {\bibfnamefont {J.}~\bibnamefont
  {Gambetta}}, \bibinfo {author} {\bibfnamefont {A.}~\bibnamefont {Blais}},
  \bibinfo {author} {\bibfnamefont {D.~I.}\ \bibnamefont {Schuster}}, \bibinfo
  {author} {\bibfnamefont {A.}~\bibnamefont {Wallraff}}, \bibinfo {author}
  {\bibfnamefont {L.}~\bibnamefont {Frunzio}}, \bibinfo {author} {\bibfnamefont
  {J.}~\bibnamefont {Majer}}, \bibinfo {author} {\bibfnamefont {M.~H.}\
  \bibnamefont {Devoret}}, \bibinfo {author} {\bibfnamefont {S.~M.}\
  \bibnamefont {Girvin}}, \ and\ \bibinfo {author} {\bibfnamefont {R.~J.}\
  \bibnamefont {Schoelkopf}},\ }\href {\doibase 10.1103/PhysRevA.74.042318}
  {\bibfield  {journal} {\bibinfo  {journal} {Phys. Rev. A}\ }\textbf {\bibinfo
  {volume} {74}},\ \bibinfo {pages} {042318} (\bibinfo {year}
  {2006})}\BibitemShut {NoStop}%
\bibitem [{\citenamefont {Rigetti}\ \emph {et~al.}(2012)\citenamefont
  {Rigetti}, \citenamefont {Gambetta}, \citenamefont {Poletto}, \citenamefont
  {Plourde}, \citenamefont {Chow}, \citenamefont {C\'orcoles}, \citenamefont
  {Smolin}, \citenamefont {Merkel}, \citenamefont {Rozen}, \citenamefont
  {Keefe}, \citenamefont {Rothwell}, \citenamefont {Ketchen},\ and\
  \citenamefont {Steffen}}]{PhysRevB.86.100506}%
  \BibitemOpen
  \bibfield  {author} {\bibinfo {author} {\bibfnamefont {C.}~\bibnamefont
  {Rigetti}}, \bibinfo {author} {\bibfnamefont {J.~M.}\ \bibnamefont
  {Gambetta}}, \bibinfo {author} {\bibfnamefont {S.}~\bibnamefont {Poletto}},
  \bibinfo {author} {\bibfnamefont {B.~L.~T.}\ \bibnamefont {Plourde}},
  \bibinfo {author} {\bibfnamefont {J.~M.}\ \bibnamefont {Chow}}, \bibinfo
  {author} {\bibfnamefont {A.~D.}\ \bibnamefont {C\'orcoles}}, \bibinfo
  {author} {\bibfnamefont {J.~A.}\ \bibnamefont {Smolin}}, \bibinfo {author}
  {\bibfnamefont {S.~T.}\ \bibnamefont {Merkel}}, \bibinfo {author}
  {\bibfnamefont {J.~R.}\ \bibnamefont {Rozen}}, \bibinfo {author}
  {\bibfnamefont {G.~A.}\ \bibnamefont {Keefe}}, \bibinfo {author}
  {\bibfnamefont {M.~B.}\ \bibnamefont {Rothwell}}, \bibinfo {author}
  {\bibfnamefont {M.~B.}\ \bibnamefont {Ketchen}}, \ and\ \bibinfo {author}
  {\bibfnamefont {M.}~\bibnamefont {Steffen}},\ }\href {\doibase
  10.1103/PhysRevB.86.100506} {\bibfield  {journal} {\bibinfo  {journal} {Phys.
  Rev. B}\ }\textbf {\bibinfo {volume} {86}},\ \bibinfo {pages} {100506}
  (\bibinfo {year} {2012})}\BibitemShut {NoStop}%
\bibitem [{\citenamefont {Majer}\ \emph {et~al.}(2007)\citenamefont {Majer},
  \citenamefont {Chow}, \citenamefont {Gambetta}, \citenamefont {Koch},
  \citenamefont {Johnson}, \citenamefont {Schreier}, \citenamefont {Frunzio},
  \citenamefont {Schuster}, \citenamefont {Houck}, \citenamefont {Wallraff},
  \citenamefont {Blais}, \citenamefont {Devoret}, \citenamefont {Girvin},\ and\
  \citenamefont {Schoelkopf}}]{Majer2007}%
  \BibitemOpen
  \bibfield  {author} {\bibinfo {author} {\bibfnamefont {J.}~\bibnamefont
  {Majer}}, \bibinfo {author} {\bibfnamefont {J.~M.}\ \bibnamefont {Chow}},
  \bibinfo {author} {\bibfnamefont {J.~M.}\ \bibnamefont {Gambetta}}, \bibinfo
  {author} {\bibfnamefont {J.}~\bibnamefont {Koch}}, \bibinfo {author}
  {\bibfnamefont {B.~R.}\ \bibnamefont {Johnson}}, \bibinfo {author}
  {\bibfnamefont {J.~A.}\ \bibnamefont {Schreier}}, \bibinfo {author}
  {\bibfnamefont {L.}~\bibnamefont {Frunzio}}, \bibinfo {author} {\bibfnamefont
  {D.~I.}\ \bibnamefont {Schuster}}, \bibinfo {author} {\bibfnamefont {A.~A.}\
  \bibnamefont {Houck}}, \bibinfo {author} {\bibfnamefont {A.}~\bibnamefont
  {Wallraff}}, \bibinfo {author} {\bibfnamefont {A.}~\bibnamefont {Blais}},
  \bibinfo {author} {\bibfnamefont {M.~H.}\ \bibnamefont {Devoret}}, \bibinfo
  {author} {\bibfnamefont {S.~M.}\ \bibnamefont {Girvin}}, \ and\ \bibinfo
  {author} {\bibfnamefont {R.~J.}\ \bibnamefont {Schoelkopf}},\ }\href
  {\doibase 10.1038/nature06184} {\bibfield  {journal} {\bibinfo  {journal}
  {Nature (London)}\ }\textbf {\bibinfo {volume} {449}},\ \bibinfo {pages} {443}
  (\bibinfo {year} {2007})}\BibitemShut {NoStop}%
\bibitem [{\citenamefont {Pellizzari}\ \emph {et~al.}(1995)\citenamefont
  {Pellizzari}, \citenamefont {Gardiner}, \citenamefont {Cirac},\ and\
  \citenamefont {Zoller}}]{PhysRevLett.75.3788}%
  \BibitemOpen
  \bibfield  {author} {\bibinfo {author} {\bibfnamefont {T.}~\bibnamefont
  {Pellizzari}}, \bibinfo {author} {\bibfnamefont {S.~A.}\ \bibnamefont
  {Gardiner}}, \bibinfo {author} {\bibfnamefont {J.~I.}\ \bibnamefont {Cirac}},
  \ and\ \bibinfo {author} {\bibfnamefont {P.}~\bibnamefont {Zoller}},\ }\href
  {\doibase 10.1103/PhysRevLett.75.3788} {\bibfield  {journal} {\bibinfo
  {journal} {Phys. Rev. Lett.}\ }\textbf {\bibinfo {volume} {75}},\ \bibinfo
  {pages} {3788} (\bibinfo {year} {1995})}\BibitemShut {NoStop}%
\bibitem [{\citenamefont {Goppl}\ \emph {et~al.}(2008)\citenamefont {Göppl},
  \citenamefont {Fragner}, \citenamefont {Baur}, \citenamefont {Bianchetti},
  \citenamefont {Filipp}, \citenamefont {Fink}, \citenamefont {Leek},
  \citenamefont {Puebla}, \citenamefont {Steffen},\ and\ \citenamefont
  {Wallraff}}]{:/content/aip/journal/jap/104/11/10.1063/1.3010859}%
  \BibitemOpen
  \bibfield  {author} {\bibinfo {author} {\bibfnamefont {M.}~\bibnamefont
  {G{\"o}ppl}}, \bibinfo {author} {\bibfnamefont {A.}~\bibnamefont {Fragner}},
  \bibinfo {author} {\bibfnamefont {M.}~\bibnamefont {Baur}}, \bibinfo {author}
  {\bibfnamefont {R.}~\bibnamefont {Bianchetti}}, \bibinfo {author}
  {\bibfnamefont {S.}~\bibnamefont {Filipp}}, \bibinfo {author} {\bibfnamefont
  {J.~M.}\ \bibnamefont {Fink}}, \bibinfo {author} {\bibfnamefont {P.~J.}\
  \bibnamefont {Leek}}, \bibinfo {author} {\bibfnamefont {G.}~\bibnamefont
  {Puebla}}, \bibinfo {author} {\bibfnamefont {L.}~\bibnamefont {Steffen}}, \
  and\ \bibinfo {author} {\bibfnamefont {A.}~\bibnamefont {Wallraff}},\ }\href
  {\doibase http://dx.doi.org/10.1063/1.3010859} {\bibfield  {journal}
  {\bibinfo  {journal} {J.Appl. Phys.}\ }\textbf {\bibinfo
  {volume} {104}},\ \bibinfo {eid} {113904} (\bibinfo {year}
  {2008})}\BibitemShut {NoStop}%
\bibitem [{\citenamefont {Benhelm}\ \emph {et~al.}(2008)\citenamefont
  {Benhelm}, \citenamefont {Kirchmair}, \citenamefont {Roos},\ and\
  \citenamefont {Blatt}}]{BenhelmIonsQIP2008}%
  \BibitemOpen
  \bibfield  {author} {\bibinfo {author} {\bibfnamefont {J.}~\bibnamefont
  {Benhelm}}, \bibinfo {author} {\bibfnamefont {G.}~\bibnamefont {Kirchmair}},
  \bibinfo {author} {\bibfnamefont {C.~F.}\ \bibnamefont {Roos}}, \ and\
  \bibinfo {author} {\bibfnamefont {R.}~\bibnamefont {Blatt}},\ }\href
  {\doibase 10.1103/PhysRevA.77.062306} {\bibfield  {journal} {\bibinfo
  {journal} {Phys. Rev. A}\ }\textbf {\bibinfo {volume} {77}},\ \bibinfo
  {pages} {062306} (\bibinfo {year} {2008})}\BibitemShut {NoStop}%
\bibitem [{\citenamefont {Barenco}\ \emph {et~al.}(1995)\citenamefont
  {Barenco}, \citenamefont {Bennett}, \citenamefont {Cleve}, \citenamefont
  {DiVincenzo}, \citenamefont {Margolus}, \citenamefont {Shor}, \citenamefont
  {Sleator}, \citenamefont {Smolin},\ and\ \citenamefont
  {Weinfurter}}]{PhysRevA.52.3457}%
  \BibitemOpen
  \bibfield  {author} {\bibinfo {author} {\bibfnamefont {A.}~\bibnamefont
  {Barenco}}, \bibinfo {author} {\bibfnamefont {C.~H.}\ \bibnamefont
  {Bennett}}, \bibinfo {author} {\bibfnamefont {R.}~\bibnamefont {Cleve}},
  \bibinfo {author} {\bibfnamefont {D.~P.}\ \bibnamefont {DiVincenzo}},
  \bibinfo {author} {\bibfnamefont {N.}~\bibnamefont {Margolus}}, \bibinfo
  {author} {\bibfnamefont {P.}~\bibnamefont {Shor}}, \bibinfo {author}
  {\bibfnamefont {T.}~\bibnamefont {Sleator}}, \bibinfo {author} {\bibfnamefont
  {J.~A.}\ \bibnamefont {Smolin}}, \ and\ \bibinfo {author} {\bibfnamefont
  {H.}~\bibnamefont {Weinfurter}},\ }\href {\doibase 10.1103/PhysRevA.52.3457}
  {\bibfield  {journal} {\bibinfo  {journal} {Phys. Rev. A}\ }\textbf {\bibinfo
  {volume} {52}},\ \bibinfo {pages} {3457} (\bibinfo {year}
  {1995})}\BibitemShut {NoStop}%
\bibitem [{\citenamefont {Lutchyn}\ \emph {et~al.}(2010)\citenamefont
  {Lutchyn}, \citenamefont {Sau},\ and\ \citenamefont
  {Das~Sarma}}]{lutchyn2010majorana}%
  \BibitemOpen
  \bibfield  {author} {\bibinfo {author} {\bibfnamefont {R.~M.}\ \bibnamefont
  {Lutchyn}}, \bibinfo {author} {\bibfnamefont {J.~D.}\ \bibnamefont {Sau}}, \
  and\ \bibinfo {author} {\bibfnamefont {S.}~\bibnamefont {Das~Sarma}},\
  }\href@noop {} {\bibfield  {journal} {\bibinfo  {journal} {Phys. Rev. Lett.}\
  }\textbf {\bibinfo {volume} {105}},\ \bibinfo {pages} {077001} (\bibinfo
  {year} {2010})}\BibitemShut {NoStop}%
\bibitem [{\citenamefont {Oreg}\ \emph {et~al.}(2010)\citenamefont {Oreg},
  \citenamefont {Refael},\ and\ \citenamefont {von Oppen}}]{oreg2010helical}%
  \BibitemOpen
  \bibfield  {author} {\bibinfo {author} {\bibfnamefont {Y.}~\bibnamefont
  {Oreg}}, \bibinfo {author} {\bibfnamefont {G.}~\bibnamefont {Refael}}, \ and\
  \bibinfo {author} {\bibfnamefont {F.}~\bibnamefont {von Oppen}},\ }\href@noop
  {} {\bibfield  {journal} {\bibinfo  {journal} {Phys. Rev. Lett.}\ }\textbf
  {\bibinfo {volume} {105}},\ \bibinfo {pages} {177002} (\bibinfo {year}
  {2010})}\BibitemShut {NoStop}%
\bibitem [{\citenamefont {Nadj-Perge}\ \emph {et~al.}(2014)\citenamefont
  {Nadj-Perge}, \citenamefont {Drozdov}, \citenamefont {Li}, \citenamefont
  {Chen}, \citenamefont {Jeon}, \citenamefont {Seo}, \citenamefont {MacDonald},
  \citenamefont {Bernevig},\ and\ \citenamefont
  {Yazdani}}]{Nadj-Perge02102014}%
  \BibitemOpen
  \bibfield  {author} {\bibinfo {author} {\bibfnamefont {S.}~\bibnamefont
  {Nadj-Perge}}, \bibinfo {author} {\bibfnamefont {I.~K.}\ \bibnamefont
  {Drozdov}}, \bibinfo {author} {\bibfnamefont {J.}~\bibnamefont {Li}},
  \bibinfo {author} {\bibfnamefont {H.}~\bibnamefont {Chen}}, \bibinfo {author}
  {\bibfnamefont {S.}~\bibnamefont {Jeon}}, \bibinfo {author} {\bibfnamefont
  {J.}~\bibnamefont {Seo}}, \bibinfo {author} {\bibfnamefont {A.~H.}\
  \bibnamefont {MacDonald}}, \bibinfo {author} {\bibfnamefont {B.~A.}\
  \bibnamefont {Bernevig}}, \ and\ \bibinfo {author} {\bibfnamefont
  {A.}~\bibnamefont {Yazdani}},\ }\href {\doibase 10.1126/science.1259327}
  {\bibfield  {journal} {\bibinfo  {journal} {Science}\ }\textbf {\bibinfo {volume} {346}},\ \bibinfo {pages} {602} (\bibinfo {year}
  {2014}).}\BibitemShut {NoStop}%
\bibitem [{\citenamefont {Fu}(2010)}]{fu2010electron}%
  \BibitemOpen
  \bibfield  {author} {\bibinfo {author} {\bibfnamefont {L.}~\bibnamefont
  {Fu}},\ }\href@noop {} {\bibfield  {journal} {\bibinfo  {journal} {Phys. Rev.
  Lett.}\ }\textbf {\bibinfo {volume} {104}},\ \bibinfo {pages} {056402}
  (\bibinfo {year} {2010})}\BibitemShut {NoStop}%
\bibitem [{\citenamefont {Kwon}\ \emph {et~al.}(2004)\citenamefont {Kwon},
  \citenamefont {Sengupta},\ and\ \citenamefont {Yakovenko}}]{Eur.Phys.37.3}%
  \BibitemOpen
  \bibfield  {author} {\bibinfo {author} {\bibfnamefont {H.-J.}\ \bibnamefont
  {Kwon}}, \bibinfo {author} {\bibfnamefont {K.}~\bibnamefont {Sengupta}}, \
  and\ \bibinfo {author} {\bibfnamefont {V.}~\bibnamefont {Yakovenko}},\ }\href
  {\doibase 10.1140/epjb/e2004-00066-4} {\bibfield  {journal} {\bibinfo
  {journal} {Europhys. J. B}\ }\textbf {\bibinfo {volume} {37}},\ \bibinfo {pages} {349}
  (\bibinfo {year} {2004})}\BibitemShut {NoStop}%
\bibitem [{\citenamefont {Wallraff}\ \emph {et~al.}(2004)\citenamefont
  {Wallraff}, \citenamefont {Schuster}, \citenamefont {Blais}, \citenamefont
  {Frunzio}, \citenamefont {Huang}, \citenamefont {Majer}, \citenamefont
  {Kumar}, \citenamefont {Girvin},\ and\ \citenamefont
  {Schoelkopf}}]{Wallraff2004}%
  \BibitemOpen
  \bibfield  {author} {\bibinfo {author} {\bibfnamefont {A.}~\bibnamefont
  {Wallraff}}, \bibinfo {author} {\bibfnamefont {D.~I.}\ \bibnamefont
  {Schuster}}, \bibinfo {author} {\bibfnamefont {A.}~\bibnamefont {Blais}},
  \bibinfo {author} {\bibfnamefont {L.}~\bibnamefont {Frunzio}}, \bibinfo
  {author} {\bibfnamefont {R.-.~S.}\ \bibnamefont {Huang}}, \bibinfo {author}
  {\bibfnamefont {J.}~\bibnamefont {Majer}}, \bibinfo {author} {\bibfnamefont
  {S.}~\bibnamefont {Kumar}}, \bibinfo {author} {\bibfnamefont {S.~M.}\
  \bibnamefont {Girvin}}, \ and\ \bibinfo {author} {\bibfnamefont {R.~J.}\
  \bibnamefont {Schoelkopf}},\ }\href {\doibase 10.1038/nature02851} {\bibfield
   {journal} {\bibinfo  {journal} {Nature (London)}\ }\textbf {\bibinfo {volume}
  {431}},\ \bibinfo {pages} {162} (\bibinfo {year} {2004})}\BibitemShut
  {NoStop}%
\bibitem [{\citenamefont {Frunzio}\ \emph {et~al.}(2005)\citenamefont
  {Frunzio}, \citenamefont {Wallraff}, \citenamefont {Schuster}, \citenamefont
  {Majer},\ and\ \citenamefont {Schoelkopf}}]{1439774}%
  \BibitemOpen
  \bibfield  {author} {\bibinfo {author} {\bibfnamefont {L.}~\bibnamefont
  {Frunzio}}, \bibinfo {author} {\bibfnamefont {A.}~\bibnamefont {Wallraff}},
  \bibinfo {author} {\bibfnamefont {D.}~\bibnamefont {Schuster}}, \bibinfo
  {author} {\bibfnamefont {J.}~\bibnamefont {Majer}}, \ and\ \bibinfo {author}
  {\bibfnamefont {R.}~\bibnamefont {Schoelkopf}},\ }\href {\doibase
  10.1109/TASC.2005.850084} {\bibfield  {journal} {\bibinfo  {journal} {IEEE Trans. Appl. Superconduct.}\ }\textbf {\bibinfo {volume} {15}},\
  \bibinfo {pages} {860} (\bibinfo {year} {2005})}\BibitemShut {NoStop}%
\bibitem [{\citenamefont {Geerlings}\ \emph {et~al.}(2012)\citenamefont
  {Geerlings}, \citenamefont {Shankar}, \citenamefont {Edwards}, \citenamefont
  {Frunzio}, \citenamefont {Schoelkopf},\ and\ \citenamefont
  {Devoret}}]{:/content/aip/journal/apl/100/19/10.1063/1.4710520}%
  \BibitemOpen
  \bibfield  {author} {\bibinfo {author} {\bibfnamefont {K.}~\bibnamefont
  {Geerlings}}, \bibinfo {author} {\bibfnamefont {S.}~\bibnamefont {Shankar}},
  \bibinfo {author} {\bibfnamefont {E.}~\bibnamefont {Edwards}}, \bibinfo
  {author} {\bibfnamefont {L.}~\bibnamefont {Frunzio}}, \bibinfo {author}
  {\bibfnamefont {R.~J.}\ \bibnamefont {Schoelkopf}}, \ and\ \bibinfo {author}
  {\bibfnamefont {M.~H.}\ \bibnamefont {Devoret}},\ }\href {\doibase
  http://dx.doi.org/10.1063/1.4710520} {\bibfield  {journal} {\bibinfo
  {journal} {Appl. Phys. Lett.}\ }\textbf {\bibinfo {volume} {100}},\
  \bibinfo {eid} {192601} (\bibinfo {year} {2012})}\BibitemShut {NoStop}%
\bibitem [{\citenamefont {Niemczyk}\ \emph {et~al.}(2010)\citenamefont
  {Niemczyk}, \citenamefont {Deppe}, \citenamefont {Huebl}, \citenamefont
  {Menzel}, \citenamefont {Hocke}, \citenamefont {Schwarz}, \citenamefont
  {Garcia-Ripoll}, \citenamefont {Zueco}, \citenamefont {Hummer}, \citenamefont
  {Solano}, \citenamefont {Marx},\ and\ \citenamefont {Gross}}]{Niemczyk2010}%
  \BibitemOpen
  \bibfield  {author} {\bibinfo {author} {\bibfnamefont {T.}~\bibnamefont
  {Niemczyk}}, \bibinfo {author} {\bibfnamefont {F.}~\bibnamefont {Deppe}},
  \bibinfo {author} {\bibfnamefont {H.}~\bibnamefont {Huebl}}, \bibinfo
  {author} {\bibfnamefont {E.~P.}\ \bibnamefont {Menzel}}, \bibinfo {author}
  {\bibfnamefont {F.}~\bibnamefont {Hocke}}, \bibinfo {author} {\bibfnamefont
  {M.~J.}\ \bibnamefont {Schwarz}}, \bibinfo {author} {\bibfnamefont {J.~J.}\
  \bibnamefont {Garcia-Ripoll}}, \bibinfo {author} {\bibfnamefont
  {D.}~\bibnamefont {Zueco}}, \bibinfo {author} {\bibfnamefont
  {T.}~\bibnamefont {Hummer}}, \bibinfo {author} {\bibfnamefont
  {E.}~\bibnamefont {Solano}}, \bibinfo {author} {\bibfnamefont
  {A.}~\bibnamefont {Marx}}, \ and\ \bibinfo {author} {\bibfnamefont
  {R.}~\bibnamefont {Gross}},\ }\href {\doibase 10.1038/nphys1730} {\bibfield
  {journal} {\bibinfo  {journal} {Nat. Phys.}\ }\textbf {\bibinfo {volume} {6}},\
  \bibinfo {pages} {772} (\bibinfo {year} {2010})}\BibitemShut {NoStop}%
\bibitem [{\citenamefont {Zueco}\ \emph {et~al.}(2009)\citenamefont {Zueco},
  \citenamefont {Reuther}, \citenamefont {Kohler},\ and\ \citenamefont
  {H\"anggi}}]{PhysRevA.80.033846}%
  \BibitemOpen
  \bibfield  {author} {\bibinfo {author} {\bibfnamefont {D.}~\bibnamefont
  {Zueco}}, \bibinfo {author} {\bibfnamefont {G.~M.}\ \bibnamefont {Reuther}},
  \bibinfo {author} {\bibfnamefont {S.}~\bibnamefont {Kohler}}, \ and\ \bibinfo
  {author} {\bibfnamefont {P.}~\bibnamefont {H\"anggi}},\ }\href {\doibase
  10.1103/PhysRevA.80.033846} {\bibfield  {journal} {\bibinfo  {journal} {Phys.
  Rev. A}\ }\textbf {\bibinfo {volume} {80}},\ \bibinfo {pages} {033846}
  (\bibinfo {year} {2009})}\BibitemShut {NoStop}%
\bibitem [{\citenamefont {Ashhab}\ and\ \citenamefont
  {Nori}(2010)}]{PhysRevA.81.042311}%
  \BibitemOpen
  \bibfield  {author} {\bibinfo {author} {\bibfnamefont {S.}~\bibnamefont
  {Ashhab}}\ and\ \bibinfo {author} {\bibfnamefont {F.}~\bibnamefont {Nori}},\
  }\href {\doibase 10.1103/PhysRevA.81.042311} {\bibfield  {journal} {\bibinfo
  {journal} {Phys. Rev. A}\ }\textbf {\bibinfo {volume} {81}},\ \bibinfo
  {pages} {042311} (\bibinfo {year} {2010})}\BibitemShut {NoStop}%
\bibitem [{\citenamefont {Leek}\ \emph {et~al.}(2010)\citenamefont {Leek},
  \citenamefont {Baur}, \citenamefont {Fink}, \citenamefont {Bianchetti},
  \citenamefont {Steffen}, \citenamefont {Filipp},\ and\ \citenamefont
  {Wallraff}}]{PhysRevLett.104.100504}%
  \BibitemOpen
  \bibfield  {author} {\bibinfo {author} {\bibfnamefont {P.~J.}\ \bibnamefont
  {Leek}}, \bibinfo {author} {\bibfnamefont {M.}~\bibnamefont {Baur}}, \bibinfo
  {author} {\bibfnamefont {J.~M.}\ \bibnamefont {Fink}}, \bibinfo {author}
  {\bibfnamefont {R.}~\bibnamefont {Bianchetti}}, \bibinfo {author}
  {\bibfnamefont {L.}~\bibnamefont {Steffen}}, \bibinfo {author} {\bibfnamefont
  {S.}~\bibnamefont {Filipp}}, \ and\ \bibinfo {author} {\bibfnamefont
  {A.}~\bibnamefont {Wallraff}},\ }\href {\doibase
  10.1103/PhysRevLett.104.100504} {\bibfield  {journal} {\bibinfo  {journal}
  {Phys. Rev. Lett.}\ }\textbf {\bibinfo {volume} {104}},\ \bibinfo {pages}
  {100504} (\bibinfo {year} {2010})}\BibitemShut {NoStop}%
\bibitem [{\citenamefont {Schuster}\ \emph {et~al.}(2005)\citenamefont
  {Schuster}, \citenamefont {Wallraff}, \citenamefont {Blais}, \citenamefont
  {Frunzio}, \citenamefont {Huang}, \citenamefont {Majer}, \citenamefont
  {Girvin},\ and\ \citenamefont {Schoelkopf}}]{PhysRevLett.94.123602}%
  \BibitemOpen
  \bibfield  {author} {\bibinfo {author} {\bibfnamefont {D.~I.}\ \bibnamefont
  {Schuster}}, \bibinfo {author} {\bibfnamefont {A.}~\bibnamefont {Wallraff}},
  \bibinfo {author} {\bibfnamefont {A.}~\bibnamefont {Blais}}, \bibinfo
  {author} {\bibfnamefont {L.}~\bibnamefont {Frunzio}}, \bibinfo {author}
  {\bibfnamefont {R.-S.}\ \bibnamefont {Huang}}, \bibinfo {author}
  {\bibfnamefont {J.}~\bibnamefont {Majer}}, \bibinfo {author} {\bibfnamefont
  {S.~M.}\ \bibnamefont {Girvin}}, \ and\ \bibinfo {author} {\bibfnamefont
  {R.~J.}\ \bibnamefont {Schoelkopf}},\ }\href {\doibase
  10.1103/PhysRevLett.94.123602} {\bibfield  {journal} {\bibinfo  {journal}
  {Phys. Rev. Lett.}\ }\textbf {\bibinfo {volume} {94}},\ \bibinfo {pages}
  {123602} (\bibinfo {year} {2005})}\BibitemShut {NoStop}%
\bibitem [{\citenamefont {Gambetta}\ \emph {et~al.}(2007)\citenamefont
  {Gambetta}, \citenamefont {Braff}, \citenamefont {Wallraff}, \citenamefont
  {Girvin},\ and\ \citenamefont {Schoelkopf}}]{PhysRevA.76.012325}%
  \BibitemOpen
  \bibfield  {author} {\bibinfo {author} {\bibfnamefont {J.}~\bibnamefont
  {Gambetta}}, \bibinfo {author} {\bibfnamefont {W.~A.}\ \bibnamefont {Braff}},
  \bibinfo {author} {\bibfnamefont {A.}~\bibnamefont {Wallraff}}, \bibinfo
  {author} {\bibfnamefont {S.~M.}\ \bibnamefont {Girvin}}, \ and\ \bibinfo
  {author} {\bibfnamefont {R.~J.}\ \bibnamefont {Schoelkopf}},\ }\href
  {\doibase 10.1103/PhysRevA.76.012325} {\bibfield  {journal} {\bibinfo
  {journal} {Phys. Rev. A}\ }\textbf {\bibinfo {volume} {76}},\ \bibinfo
  {pages} {012325} (\bibinfo {year} {2007})}\BibitemShut {NoStop}%
\bibitem [{\citenamefont {Johnson}\ \emph {et~al.}(2010)\citenamefont
  {Johnson}, \citenamefont {Reed}, \citenamefont {Houck}, \citenamefont
  {Schuster}, \citenamefont {Bishop}, \citenamefont {Ginossar}, \citenamefont
  {Gambetta}, \citenamefont {DiCarlo}, \citenamefont {Frunzio}, \citenamefont
  {Girvin},\ and\ \citenamefont {Schoelkopf}}]{Johnson2010}%
  \BibitemOpen
  \bibfield  {author} {\bibinfo {author} {\bibfnamefont {B.~R.}\ \bibnamefont
  {Johnson}}, \bibinfo {author} {\bibfnamefont {M.~D.}\ \bibnamefont {Reed}},
  \bibinfo {author} {\bibfnamefont {A.~A.}\ \bibnamefont {Houck}}, \bibinfo
  {author} {\bibfnamefont {D.~I.}\ \bibnamefont {Schuster}}, \bibinfo {author}
  {\bibfnamefont {L.~S.}\ \bibnamefont {Bishop}}, \bibinfo {author}
  {\bibfnamefont {E.}~\bibnamefont {Ginossar}}, \bibinfo {author}
  {\bibfnamefont {J.~M.}\ \bibnamefont {Gambetta}}, \bibinfo {author}
  {\bibfnamefont {L.}~\bibnamefont {DiCarlo}}, \bibinfo {author} {\bibfnamefont
  {L.}~\bibnamefont {Frunzio}}, \bibinfo {author} {\bibfnamefont {S.~M.}\
  \bibnamefont {Girvin}}, \ and\ \bibinfo {author} {\bibfnamefont {R.~J.}\
  \bibnamefont {Schoelkopf}},\ }\href {\doibase 10.1038/nphys1710} {\bibfield
  {journal} {\bibinfo  {journal} {Nat. Phys.}\ }\textbf {\bibinfo {volume} {6}},\
  \bibinfo {pages} {663} (\bibinfo {year} {2010})}\BibitemShut {NoStop}%
\bibitem [{\citenamefont {Fragner}\ \emph {et~al.}(2008)\citenamefont
  {Fragner}, \citenamefont {Göppl}, \citenamefont {Fink}, \citenamefont
  {Baur}, \citenamefont {Bianchetti}, \citenamefont {Leek}, \citenamefont
  {Blais},\ and\ \citenamefont {Wallraff}}]{Fragner28112008}%
  \BibitemOpen
  \bibfield  {author} {\bibinfo {author} {\bibfnamefont {A.}~\bibnamefont
  {Fragner}}, \bibinfo {author} {\bibfnamefont {M.}~\bibnamefont {G{\"o}ppl}},
  \bibinfo {author} {\bibfnamefont {J.~M.}\ \bibnamefont {Fink}}, \bibinfo
  {author} {\bibfnamefont {M.}~\bibnamefont {Baur}}, \bibinfo {author}
  {\bibfnamefont {R.}~\bibnamefont {Bianchetti}}, \bibinfo {author}
  {\bibfnamefont {P.~J.}\ \bibnamefont {Leek}}, \bibinfo {author}
  {\bibfnamefont {A.}~\bibnamefont {Blais}}, \ and\ \bibinfo {author}
  {\bibfnamefont {A.}~\bibnamefont {Wallraff}},\ }\href {\doibase
  10.1126/science.1164482} {\bibfield  {journal} {\bibinfo  {journal}
  {Science}\ }\textbf {\bibinfo {volume} {322}},\ \bibinfo {pages} {1357}
  (\bibinfo {year} {2008})}\BibitemShut {NoStop}%
\bibitem [{\citenamefont {Wallraff}\ \emph {et~al.}(2005)\citenamefont
  {Wallraff}, \citenamefont {Schuster}, \citenamefont {Blais}, \citenamefont
  {Frunzio}, \citenamefont {Majer}, \citenamefont {Devoret}, \citenamefont
  {Girvin},\ and\ \citenamefont {Schoelkopf}}]{PhysRevLett.95.060501}%
  \BibitemOpen
  \bibfield  {author} {\bibinfo {author} {\bibfnamefont {A.}~\bibnamefont
  {Wallraff}}, \bibinfo {author} {\bibfnamefont {D.~I.}\ \bibnamefont
  {Schuster}}, \bibinfo {author} {\bibfnamefont {A.}~\bibnamefont {Blais}},
  \bibinfo {author} {\bibfnamefont {L.}~\bibnamefont {Frunzio}}, \bibinfo
  {author} {\bibfnamefont {J.}~\bibnamefont {Majer}}, \bibinfo {author}
  {\bibfnamefont {M.~H.}\ \bibnamefont {Devoret}}, \bibinfo {author}
  {\bibfnamefont {S.~M.}\ \bibnamefont {Girvin}}, \ and\ \bibinfo {author}
  {\bibfnamefont {R.~J.}\ \bibnamefont {Schoelkopf}},\ }\href {\doibase
  10.1103/PhysRevLett.95.060501} {\bibfield  {journal} {\bibinfo  {journal}
  {Phys. Rev. Lett.}\ }\textbf {\bibinfo {volume} {95}},\ \bibinfo {pages}
  {060501} (\bibinfo {year} {2005})}\BibitemShut {NoStop}%
\bibitem{cohen1998atom}
	C.~Cohen-Tannoudji, J.~Dupont-Roc, and G.~Grynberg.
	\newblock {\em Atom-Photon Interactions: Basic Processes and Applications}.
	\newblock (Wiley, New York, 1998).
\bibitem [{\citenamefont {{Higginbotham}}\ \emph {et~al.}(2015)\citenamefont
  {{Higginbotham}}, \citenamefont {{Albrecht}}, \citenamefont {{Kirsanskas}},
  \citenamefont {{Chang}}, \citenamefont {{Kuemmeth}}, \citenamefont
  {{Krogstrup}}, \citenamefont {{Jespersen}}, \citenamefont {{Nygard}},
  \citenamefont {{Flensberg}},\ and\ \citenamefont
  {{Marcus}}}]{2015arXiv150105155H}%
  \BibitemOpen
  \bibfield  {author} {\bibinfo {author} {\bibfnamefont {A.~P.}\ \bibnamefont
  {{Higginbotham}}}, \bibinfo {author} {\bibfnamefont {S.~M.}\ \bibnamefont
  {{Albrecht}}}, \bibinfo {author} {\bibfnamefont {G.}~\bibnamefont
  {{Kirsanskas}}}, \bibinfo {author} {\bibfnamefont {W.}~\bibnamefont
  {{Chang}}}, \bibinfo {author} {\bibfnamefont {F.}~\bibnamefont {{Kuemmeth}}},
  \bibinfo {author} {\bibfnamefont {P.}~\bibnamefont {{Krogstrup}}}, \bibinfo
  {author} {\bibfnamefont {T.~S.}\ \bibnamefont {{Jespersen}}}, \bibinfo
  {author} {\bibfnamefont {J.}~\bibnamefont {{Nygard}}}, \bibinfo {author}
  {\bibfnamefont {K.}~\bibnamefont {{Flensberg}}}, \ and\ \bibinfo {author}
  {\bibfnamefont {C.~M.}\ \bibnamefont {{Marcus}}},\ }\href@noop {} {\bibfield
  {journal} (\bibinfo {year}
  {2015})},\ \Eprint {http://arxiv.org/abs/1501.05155} {arXiv:1501.05155} \BibitemShut {NoStop}%
\bibitem [{\citenamefont {Goldstein}(1929)}]{1929goldstein}%
  \BibitemOpen
  \bibfield  {author} {\bibinfo {author} {\bibfnamefont {S.}~\bibnamefont
  {Goldstein}},\ }\href@noop {} {\bibfield  {journal} {\bibinfo  {journal}
  {Proc. R. Soc. Edinburgh}\ }\textbf {\bibinfo {volume} {49}},\ \bibinfo
  {pages} {210} (\bibinfo {year} {1930})}\BibitemShut {NoStop}%
\bibitem [{\citenamefont {{Crothers}}(1976)}]{1976JPhB....9L.513C}%
  \BibitemOpen
  \bibfield  {author} {\bibinfo {author} {\bibfnamefont {D.~S.~F.}\
  \bibnamefont {{Crothers}}},\ }\href {\doibase 10.1088/0022-3700/9/17/001}
  {\bibfield  {journal} {\bibinfo  {journal} {J. Phys. B}\ }\textbf {\bibinfo {volume} {9}},\ \bibinfo {pages}
  {L513} (\bibinfo {year} {1976})}\BibitemShut {NoStop}%
\bibitem [{\citenamefont {Jeffreys}(1923)}]{1923jeffreys}%
  \BibitemOpen
  \bibfield  {author} {\bibinfo {author} {\bibfnamefont {H.}~\bibnamefont
  {Jeffreys}},\ }\href@noop {} {\bibfield  {journal} {\bibinfo  {journal}
  {Proc. London Math. Soc.}\ }\textbf {\bibinfo {volume} {s2-23}},\ \bibinfo
  {pages} {437} (\bibinfo {year} {1925})}\BibitemShut {NoStop}%
\bibitem [{\citenamefont {Abramowitz}\ and\ \citenamefont
  {Stegun}(1970)}]{abramowitz}%
  \BibitemOpen
  \bibfield  {author} {\bibinfo {author} {\bibfnamefont {M.}~\bibnamefont
  {Abramowitz}}\ and\ \bibinfo {author} {\bibfnamefont {I.~A.}\ \bibnamefont
  {Stegun}},\ }\href@noop {} {\emph {\bibinfo {title} {Handbook of Mathematical
  Functions}}}\ (\bibinfo  {publisher} {Dover},\ \bibinfo
  {address} {New York},\ \bibinfo {year} {1970})\BibitemShut {NoStop}%
\end{thebibliography}
\end{document}